\documentclass[12pt]{article}

\usepackage{amsmath,amssymb,amsfonts,graphics,graphicx,amscd,amsfonts,epsfig,color}
\usepackage{subfig}
\usepackage{cite}
\usepackage{enumitem}
\usepackage{here}

\newcommand {\beq} {\begin{equation}}
\newcommand {\eeq} {\end{equation}}
\newcommand {\beqa}{\begin{eqnarray}}
\newcommand {\eeqa}{\end{eqnarray}}

\newcommand {\tr}{{\rm tr\,}}
\newcommand {\Tr}{\mbox{Tr\,}}

\newcommand {\ee}{\mbox{e}}

\makeatletter
    
    \@addtoreset{equation}{section}
  \makeatother
\allowdisplaybreaks[4]

\date{}
\begin{document}

\begin{flushright} 
%\today\\
% WIS/06/09-MAY-DPP\\ 
KEK-TH-2110
\end{flushright} 

\vspace{0.1cm}

\begin{center}
%{\LARGE The mechanism for the emergence of (3+1)D expanding} \\[1mm]
%{\LARGE space-time in the Lorentzian type IIB matrix model}
%% {\LARGE On the emergence of (3+1)-dimensional expanding} \\[1mm]
%% {\LARGE space-time in the Lorentzian type IIB matrix model}
{\LARGE On the structure of the emergent 3d expanding space} \\[1mm]
{\LARGE in the Lorentzian type IIB matrix model}
%\\[1mm]
%{\LARGE at finite density and low temperature} \\[1mm]
%{\LARGE with the deformation technique}
\end{center}

\vspace{0.1cm}
\vspace{0.1cm}
\begin{center}

%%          Keitaro N{\sc agata}$^{ab}$\footnote
%%           {
%%  E-mail address : k-nagata@kochi-u.ac.jp}, 
         Toshihiro A{\sc oki}$^{a}$\footnote
          {
 E-mail address : toshaoki@post.kek.jp}, 
         Mitsuaki H{\sc irasawa}$^{a}$\footnote
          {
 E-mail address : mitsuaki@post.kek.jp}, 
         Yuta I{\sc to}$^{b}$\footnote
          {
 E-mail address : yito@post.kek.jp}, 

         Jun N{\sc ishimura}$^{ab}$\footnote
          {
 E-mail address : jnishi@post.kek.jp} 
and
         Asato T{\sc suchiya}$^{c}$\footnote
          {
 E-mail address : tsuchiya.asato@shizuoka.ac.jp} 

\vspace{0.5cm}

$^a${\it Department of Particle and Nuclear Physics, 
School of High Energy Accelerator Science,\\
Graduate University for Advanced Studies (SOKENDAI),\\
1-1 Oho, Tsukuba, Ibaraki 305-0801, Japan} 

$^b${\it KEK Theory Center, 
High Energy Accelerator Research Organization,\\
1-1 Oho, Tsukuba, Ibaraki 305-0801, Japan} 

$^c${\it Department of Physics, Shizuoka University,\\
836 Ohya, Suruga-ku, Shizuoka 422-8529, Japan}

\end{center}
%\newpage

\vspace{1.5cm}

\begin{center}
  {\bf abstract}
\end{center}

\noindent 
The emergence of 
(3+1)-dimensional expanding space-time
in the Lorentzian type IIB matrix model
is an 
%fascinating
intriguing 
phenomenon which was observed 
in Monte Carlo studies of this model.
In particular, this may be taken as a support to the conjecture
that the model is a nonperturbative formulation of superstring theory
in (9+1) dimensions.
In this paper we investigate the space-time structure of the matrices
generated by simulating this model and its simplified versions,
and find that the expanding part of the space is described essentially
by the Pauli matrices.
%% While this clarifies the mechanism for the spontaneous breaking
%% of the SO(9) rotational symmetry,
%% it raises an important question
%% of how one can obtain a regular space-time from this model.
We argue that this is due to 
an approximation used in the simulation to avoid the sign problem,
which actually amounts to 
replacing $\ee^{iS_{\rm b}}$ by $\ee^{\beta S_{\rm b}}$ ($\beta>0$)
in the partition function, 
where $S_{\rm b}$ is the bosonic part of the action.
% and $\beta$ is some positive constant.
We also discuss the possibility of obtaining 
a regular space-time 
with the (3+1)-dimensional expanding behavior
in the original model with the correct $\ee^{iS_{\rm b}}$ factor.
\newpage

%%%%%%%%%%%%%%%%%%%%%%%%%%%%%%%%%%%%%%%%%%%%%%%%%%%%%%%%%%%%%%%%%%%%%%
%%%%%%%%%%%%%%%%%%%%%%%%%%%%%%%%%%%%%%%%%%%%%%%%%%%%%%%%%%%%%%%%%%%%%%
%%%%%%%%%%%%%%%%%%%%%%%%%%%%%%%%%%%%%%%%%%%%%%%%%%%%%%%%%%%%%%%%%%%%%%
\section{Introduction}
%\hspace{0.51cm}
%%%%%%%%%%%%%%%%%%%%%%%%%%%%%%%%%%%%%%%%%%%%%%%%%%%%%%%%%%%%%%%%%%%%%%
%%%%%%%%%%%%%%%%%%%%%%%%%%%%%%%%%%%%%%%%%%%%%%%%%%%%%%%%%%%%%%%%%%%%%%
%%%%%%%%%%%%%%%%%%%%%%%%%%%%%%%%%%%%%%%%%%%%%%%%%%%%%%%%%%%%%%%%%%%%%%

Superstring theory is a promising candidate for a unified theory
that includes quantum gravity consistently. One of the striking
consequences of this theory is that the space-time 
should have ten dimensions. 
Therefore, in order to 
make the theory compatible with 
%account for 
our (3+1)d world, the extra six dimensions have to be 
compactified somehow.
Depending on the structure of these compact extra dimensions, one can
obtain various quantum field theories in the (3+1)d space-time
at low energy. 
This issue has been investigated extensively
at the perturbative level including D-branes
configurations as a background accounting for certain nonperturbative 
effects, and it led to tremendously many consistent vacua, 
the situation which is called the landscape.
However, it remains to be seen what really happens 
if one formulates the theory in a fully nonperturbative manner
as one does in the case of quantum field theory 
using the lattice formulation.
%in a fully 
%nonpertubative formulation of superstring theory.

%Back in 1996, 
The type IIB matrix model \cite{Ishibashi:1996xs} was proposed
as such a nonperturbative formulation of superstring theory.
Formally, the action of the model can be obtained by
dimensionally reducing the action of 10d $\mathcal{N}=1$ SYM theory to 0d,
and it actually has maximal $\mathcal{N}=2$ supersymmetry in 10d.
The space-time does not exist \emph{a priori}, 
and it is represented by the eigenvalue distribution of 
the ten bosonic matrices $A_\mu$ $(\mu=0, \cdots ,9)$.
This is manifested by the fact that
%Indeed 
translations in the supersymmetry algebra turn out to be realized by
the shifts $A_\mu \mapsto A_\mu + \alpha_\mu {\bf 1}$ 
in the ten bosonic matrices.
%appears from the matrix degrees of freedom 
%This implies that the model 
%type IIB matrix
The model, therefore,
has the potential to clarify a possible nonperturbative mechanism
for dynamical compactification in superstring theory.
The Euclidean version of the model
was investigated from this viewpoint,
and the spontaneous breaking of SO(10) rotational symmetry 
was suggested by various 
approaches \cite{Nishimura:2001sx,Kawai:2002jk,Aoyama:2006rk,Nishimura:2011xy,Anagnostopoulos:2013xga,Anagnostopoulos:2017gos}.
However, latest calculations based on the Gaussian expansion method
suggested that the SO(10) symmetry is broken down to SO(3) 
instead of SO(4) \cite{Nishimura:2011xy}.
%hthe rotational symmetry is broken down to 
%% Especially, the recent study investigating the spontaneous symmetry
%% breaking in the Euclidean type IIB matrix model \cite{}
%% suggests that the vacuum is more likely to be three dimensions rather
%% than four dimensions.

This provided a strong motivation to consider
the Lorentzian version of the model.
%motivated Monte Carlo studies
%of the Lorentzian type IIB matrix model \cite{Kim:2011cr},
Monte Carlo simulation was performed in ref.~\cite{Kim:2011cr},
and the results turned out to be intriguing.
%revealed interesting properties.
In the SU($N$) basis which diagonalizes the temporal matrix $A_0$,
the spatial matrices $A_i$ $(i=1,\cdots ,9)$
have a band-diagonal structure, which
enabled the extraction of the real-time evolution.
%The extent of space at each time is then probed by 
In this way, it was found that only three out of 
nine directions start to expand
at some critical time, which implies that the model predicts
the emergence of a (3+1)d expanding space-time 
from superstring theory in (9+1)d.
The expanding behavior for a longer time was 
investigated by simulating simplified models.
%of the type IIB matrix model.
The obtained results suggested a scenario for the full model 
that the expansion is exponential
at early times \cite{Ito:2013ywa},
which is reminiscent of the inflation,
and that it turns into a power law \cite{Ito:2015mxa}
at later times, which is reminiscent of
the Friedmann-Robertson-Walker universe in the radiation dominated era.
See also refs.~\cite{Yang:2015vna,Kim:2018mfv,Tomita:2015let}
for closely related work.
%% The obtained results suggested a scenario for the full model 
%% that the expansion is exponential
%% at early times\footnote{This 
%% behavior was observed
%% also in the original type IIB matrix model \cite{Ito:2013qga} although 
%% the matrix size used was not large enough to confirm the long-time 
%% behavior.} \cite{Ito:2013ywa},
%% and turns into a power law \cite{Ito:2015mxa}
%% at later times
%% as suggested by the cosmic inflation and
%% the subsequent transition into the radiation dominated era
%% in the Friedmann-Robertson-Walker universe.

Due to the expansion of space, it is expected that
the dominant configurations can be well approximated 
at late times
by some classical solution 
of the Lorentzian type IIB matrix model.
Indeed several types of classical solutions representing
expanding space-time have been 
constructed \cite{Kim:2011ts,Kim:2012mw,Chaney:2015ktw,Chaney:2015mfa,Chaney:2016npa,Stern:2018wud,Steinacker:2017vqw,Steinacker:2017bhb}.
Also, matrix configurations with various structures in the extra dimensions 
are considered to realize chiral fermions in the (3+1)d 
space-time.
Earlier attempts used slightly modified models obtained, for instance, 
by orbifolding \cite{Aoki:2002jt,Chatzistavrakidis:2010xi}
or by toroidal compactification with magnetic fluxes \cite{Aoki:2010gv}.
More recently \cite{Chatzistavrakidis:2011gs,Nishimura:2013moa}, 
it was shown that the original model can be used
to realize 
the idea of intersecting D-branes \cite{Berkooz:1996km},
which led to the proposal of
matrix configurations that can give rise to phenomenologically viable
low-energy effective 
theories \cite{Steinacker:2014fja,Aoki:2014cya,Honda:2019bdi}.
%discussed \cite{Steinacker:2016vgf,Steinacker:2014fja,Nishimura:2013moa,}.

In this paper we investigate the space-time structure of the 
%expanding space-time
matrix configurations generated 
by Monte Carlo simulation of the Lorentzian type IIB matrix model
and the simplified models.
In particular, we calculate the eigenvalues of the submatrices 
of the spatial matrices $A_i$
corresponding to each time slice and find that only two
% of them
eigenvalues
grow in magnitude after the critical time.
A more detailed analysis shows that 
the expanding 3d space is 
described essentially by the Pauli matrices.
Namely the space is actually 
more like a fuzzy sphere
although it has been called ``3d'' in the sense that
it has three extended directions.
While we keep on using the word 3d in this sense in what follows,
we refer to the space with the Pauli-matrix structure
as a ``singular 3d space''.
%% Throughout this paper, we keep on using the word ``3d'' in this sense.
%% However, in order to remember that 
%% the space with the Pauli-matrix structure
%% is not 3d in the usual sense, we refer to it as a ``singular 3d space''.
%with a coefficient growing with time.
We observe that the situation remains unaltered even at late times or
in the continuum limit for the simplified models, and it is
shared by the original model with maximal supersymmetry as well.
%, and even in the presence of the fermionic matrices.
This raises the important question of whether
this model can generate a 3d space with continuum geometry,
which we refer to as a ``regular 3d space''. 

%%
%% While this observation confirms
%% the mechanism for the spontaneous breaking of the SO(9) rotational symmetry,
%% which was previously suggested only by a hand-waving arguement \cite{Kim:2011cr},
%% it also raises the important question of how one can obtain
%% a regular space-time from this model.
%%
%% This observation confirms
%% the speculated mechanism for the emergence of 
%% three expanding directions based on a hand-waving argument \cite{Kim:2011cr},
%% while it raises the important question of how to obtain
%% a regular space-time from this model.

In fact, Monte Carlo simulation
of the Lorentzian type IIB matrix model is not straightforward
due to the phase factor $\ee^{iS_{\rm b}}$
in the partition function,
where $S_{\rm b}$ is the bosonic part of the action.
The importance sampling is not applicable as it is and
one has to face the sign problem if one uses 
reweighting for this factor.
In this work as well as in the previous studies,
this problem is avoided by 
integrating out the scale factor of the bosonic matrices $A_\mu$ first
and using certain approximation.
%% this problem was avoided by replacing the factor $\ee^{iS_{\rm b}}$
%% by the delta function
%% $\delta (S_{\rm b})$ using certain approximation
%% after integrating out the scale factor of the bosonic matrices $A_\mu$.
Here we point out a subtlety in this approximation, and argue that 
it actually amounts to
% simulating a model 
%with a positive definite weight $\ee^{S_{\rm b}}$ instead of $\ee^{iS_{\rm b}}$.
replacing the phase factor 
$\ee^{iS_{\rm b}}$ by a positive weight $\ee^{\beta S_{\rm b}}$ 
($\beta > 0$).
This new interpretation of the simulations naturally
explains not only the emergence of the band-diagonal structure in the
spatial matrices $A_i$, which is crucial
in extracting the real-time evolution, but also the 
(3+1)d expanding behavior with the Pauli-matrix structure.
% in $A_i$, which is crucial
%in the spotaneous breaking of the SO(9) rotational symmetry.
We also discuss the possibility
of obtaining a regular space-time
in the original model with the phase factor $\ee^{iS_{\rm b}}$
without spoiling the (3+1)d expanding behavior.
Some results supporting this possibility
are reported in a separate paper \cite{workinprog},
where the sign problem is overcome by
using the complex Langevin method.
%at least for the bosonic model.

The rest of this paper is organized as follows. 
In section \ref{sec:review}
we briefly review the Lorentzian type IIB matrix model.
In section \ref{sec:space-time-structure} we discuss the
space-time structure of the matrix configurations obtained by 
simulation,
%In section \ref{sec:pauli-matrix} 
and show that they are essentially described by the Pauli matrices.
In section \ref{sec:approximation} we provide theoretical understanding
of the obtained results, and discuss the possibility 
of obtaining a regular space-time with
the (3+1)d expanding behavior
%in the Lorentzian type IIB matrix model 
if the sign problem is treated correctly.
%% argue that the approximation
%% used in Monte Carlo simulation to overcome the sign problem
%% actually amounts to replacing $\ee^{iS_{\rm b}}$ 
%% by $\ee^{\beta S_{\rm b}}$.
Section \ref{sec:summary} is devoted to a summary and discussions.

\section{Brief review of the Lorentzian type IIB matrix model}
\label{sec:review}

In this section, we define 
the Lorentzian type IIB matrix model and its simplified versions,
and review some results obtained by Monte Carlo simulations.

\subsection{Definition of the Lorentzian type IIB matrix model}

The action of the 
%Lorentzian version of 
type IIB matrix model
is given as \cite{Ishibashi:1996xs}
\begin{eqnarray}
S & = & S_{\mathrm{b}}+S_{\mathrm{f}} \ ,
\label{eq:action}\\
S_{\mathrm{b}} & = & 
%-\frac{1}{4g^{2}}\mathrm{Tr}
-\frac{1}{4}\mathrm{Tr}
\Big(\left[A_{\mu},A_{\nu}\right]\left[A^{\mu},A^{\nu}\right]\Big) \ ,
\label{eq:sb}\\
S_{\mathrm{f}} & = & 
%-\frac{1}{2g^{2}}\mathrm{Tr}
-\frac{1}{2}\mathrm{Tr}
\Big(\Psi_{\alpha}
\left(\mathcal{C}\Gamma^{\mu}\right)_{\alpha\beta}
\left[A_{\mu},\Psi_{\beta}\right]\Big) \ , 
\label{eq:sf}
\end{eqnarray}
where $A_{\mu}$ ($\mu=0,1, \cdots,9$) and $\Psi_{\alpha}$ ($\alpha=1,\cdots,16$)
are bosonic and fermionic $N\times N$ traceless Hermitian matrices.
The indices $\mu$ and $\nu$ are contracted
with the Lorentzian metric $\eta_{\mu\nu}=\mathrm{diag}(-1,1,\ldots,1)$.
The $16\times16$ matrices $\Gamma^{\mu}$ and $\mathcal{C}$
are the 10-dimensional gamma matrices and the charge conjugation matrix,
respectively,
obtained after the Weyl projection. 
The action (\ref{eq:action}) has a manifest SO(9,1)
Lorentz symmetry, under which $A_{\mu}$ and $\Psi_{\alpha}$ transform
as a Lorentz vector and a Majorana-Weyl spinor, respectively. 
%% The coupling constant $g$ is merely a scale parameter, 
%% which can be absorbed by rescaling $A_{\mu}$ and $\Psi_{\alpha}$ 
%% appropriately.

The partition function of the Lorentzian type IIB matrix model is
defined as \cite{Kim:2011cr}
\beq
Z  =  \int dA \, d\Psi\, \ee^{iS\left[A,\Psi\right]}
 =  \int dA \, \mathrm{Pf}\mathcal{M}(A)\,\ee^{iS_{\mathrm{b}}} \ ,
\label{eq:partition-function}
\eeq
%% \begin{eqnarray}
%% Z & = & \int dA \, d\Psi\, \ee^{iS\left[A,\Psi\right]}
%% \label{eq:partition-function0}
%% \\
%%  & = & \int dA \, \mathrm{Pf}\mathcal{M}(A)\,\ee^{iS_{\mathrm{b}}} \ ,
%% \label{eq:partition-function}
%% \end{eqnarray}
where the ``$i$'' in front of the action is motivated from
the fact that the string worldsheet metric has a Lorentzian
signature. Note that the bosonic action $S_{\mathrm{b}}$
%(\ref{eq:sb}) 
can be written as
\begin{equation}
S_{\mathrm{b}}=
%\frac{1}{4g^{2}}\mathrm{Tr}\left(F_{\mu\nu}F^{\mu\nu}\right)
%=\frac{1}{4g^{2}}\left\{ -2\mathrm{Tr}\left(F_{0i}\right)^{2}
\frac{1}{4}\mathrm{Tr}\left(F_{\mu\nu}F^{\mu\nu}\right)
=\frac{1}{4}\left\{ -2\mathrm{Tr}\left(F_{0i}\right)^{2}
+\mathrm{Tr}\left(F_{ij}\right)^{2}\right\} \ ,
\label{eq:sb2}
\end{equation}
where we have introduced the Hermitian matrices 
$F_{\mu\nu}=i\left[A_{\mu},A_{\nu}\right]$. 
Hence $S_{\mathrm{b}}$ is not positive semi-definite 
unlike in the Euclidean case.
%because the two terms in the last expression of eq.(\ref{eq:sb2}) have opposite signs.
Note also that, unlike in the Euclidean 
version \cite{Krauth:1998xh,Austing:2001pk},
the matrix integral in (\ref{eq:partition-function})
is divergent because $\ee^{iS_{\mathrm{b}}}$ is a pure phase factor
and the Pfaffian $\mathrm{Pf}\mathcal{M}(A)$ obtained
by integrating out the fermionic matrices is a polynomial in $A_\mu$.

In order to make the partition function (\ref{eq:partition-function}) finite,
we need to introduce the IR cutoffs 
both in the temporal and spatial directions, 
for instance, as
\begin{eqnarray}
\frac{1}{N}\mathrm{Tr}
\left\{ \left(A_{0}\right)^{2}\right\} ^{p} 
& \leq & \kappa^{p}\frac{1}{N}
\mathrm{Tr}\left\{ \left(A_{i}\right)^{2}\right\} ^{p}\ ,
\label{eq:cutoff-temp}\\
\frac{1}{N}\mathrm{Tr}\left\{ \left(A_{i}\right)^{2}\right\} ^{p} 
& \leq & L^{2p}\ .
\label{eq:cutoff-space}
\end{eqnarray}
The power $p$ is a parameter, 
which can be used to test how much the obtained results
depend on the way the IR cutoff is introduced \cite{Ito:2017rcr}.
While $p=1$ would be a natural choice,
it was proposed that $p$ should be chosen 
to be a slightly larger value
in order to make the results almost independent of $p$.
Too large values of $p$ lead to pathological behaviors, however.
%, which typically requires $p\sim 1.5$.

%one needs to choose appropriately so that the cutoff effects disappear. 
%We will explain later in this section how the value of $p$ should be chosen. 

%% The Pfaffian $\mathrm{Pf}\mathcal{M}(A)$ in 
%% (\ref{eq:partition-function}) obtained
%% by integrating out the fermion matrices
%% is real in the Lorentzian version unlike in the Euclidean version.

The Pfaffian $\mathrm{Pf}\mathcal{M}(A)$ in 
(\ref{eq:partition-function}) 
%obtained by integrating out the fermionic matrices
is real
in the Lorentzian version unlike in the Euclidean version,
where it becomes complex due to the replacement $A_0 = i A_{10}$.
However, the phase factor $\ee^{iS_{\mathrm{b}}}$ causes
the sign problem when one tries to investigate the Lorentzian model 
by Monte Carlo methods.
Here, we avoid this problem\footnote{Strictly speaking,
the model (\ref{eq:partition-function2}) is not completely
free of sign-problem 
because the Pfaffian is real but not positive semi-definite.
%can be both positive and negative. 
However, configurations with positive Pfaffian dominates
the path integral (\ref{eq:partition-function2}) at large $N$, 
and therefore one can safely
replace the Pfaffian by its absolute value in the simulation.}
following previous work \cite{Kim:2011cr,Ito:2013ywa,Ito:2015mxa}
by 
rewriting the partition function
(\ref{eq:partition-function}) 
%imposing the IR cutoffs (\ref{eq:cutoff-temp}) and (\ref{eq:cutoff-space}) 
as
\begin{align}
Z & =\int dA \, \mathrm{Pf}\mathcal{M}(A) \, 
\delta \Big(\frac{1}{N}\mathrm{Tr} F_{\mu\nu}F^{\mu\nu}-C \Big) \, 
\delta \Big(\frac{1}{N}\mathrm{Tr}
\{ (A_{i})^{2} \} ^{p}-1 \Big) \, 
\theta \Big( \kappa^{p}-\frac{1}{N}
\mathrm{Tr} \{ (A_{0})^{2}\} ^{p} \Big)  \ ,
\label{eq:partition-function2}
\end{align}
where $\theta(x)$ is the Heaviside step function.
This can be obtained by
integrating out the overall
scale factor of the bosonic matrices $A_\mu$ first
and using certain 
approximation 
as discussed in section \ref{sec:approximation}.
% (See Appendix A of Ref.~\cite{Ito:2013ywa} for the details.)
The parameter $C$ should be set to zero according to the ``derivation'', 
but
we generalize the model by choosing $C\neq 0$, which allows us to
obtain results for larger matrices in the original $C=0$ model by using
smaller matrices \cite{Azuma:2017dcb,Ito:2013ywa}.
See Appendix B of ref.~\cite{Ito:2013ywa} for the details
of the Monte Carlo simulation 
of the model (\ref{eq:partition-function2}).
% can be performed without the sign problem.
%by the standared RHMC algorithm.
%

%\subsection{Extracting the time-evolution}
\subsection{SSB of rotational SO(9) symmetry}
\label{sec:time-evolution}

Next we discuss how one can extract the time-evolution from a given matrix
configuration generated by Monte Carlo simulation \cite{Kim:2011cr}.
Since the eigenvalues of the temporal matrix $A_{0}$ represents time,
we work in an SU($N$) basis which diagonalizes $A_{0}$ as
\begin{equation}
A_{0}=\mathrm{diag}(\alpha_{1},\ldots,\alpha_{N}) \ ,
\mbox{~where~}\alpha_{1}<\cdots<\alpha_{N} \ .
\label{eq:gauge-fixing}
\end{equation}
%using the SU($N$) symmetry. 
In this basis, the spatial matrices $A_{i}$
%generated by the Monte Carlo simulations 
turn out to have an approximate
band-diagonal structure. 
By this, we mean that there exists\footnote{In practice,
the integer $n$ can be determined by
observing the scaling behavior for $\sum_i |(A_i)_{ab}|^2$
with $(a+b)/2$ fixed to different values
corresponding to different time slices.
See section 5 of ref. \cite{Ito:2015mxa} for the details.}
some integer
$n$ such that the elements of the spatial matrices $\left(A_{i}\right)_{ab}$
for $|a-b| >  n$ are much smaller than those for $|a-b|<n$. 
%% In actual simulations, this integer $n$ is determined by observing that
%% the magnitude of matrix elements $|\left(A_{i}\right)_{ab}|$ damps
%% sufficiently as one goes away from the diagonal part, which was indeed
%% performed, for instance, in section 5 of ref. \cite{Ito:2015mxa}.
%% In what follows, we assume that the spatial matrices $A_{i}$ has
%% such a band-diagonal structure with a certain integer $n<N$. 
Thanks to this structure,
we can
naturally consider the $n\times n$ submatrices $\bar{A}_{i}$
\begin{equation}
\left(\bar{A}_{i}\right)_{IJ}(t)\equiv\left(A_{i}\right)_{\nu+I,\nu+J}
\label{def-abar}
\end{equation}
representing the state
% of the 9d space 
at time $t$ defined by
\begin{equation}
t\equiv\frac{1}{n}\sum_{I=1}^{n}\alpha_{\nu+I} \ ,
\label{eq:def-time}
\end{equation}
where $I,J=1,\ldots,n$ and $\nu=0,1,\ldots,N-n$. For example, we
can define the extent of the 9d space at time $t$ using $\bar{A}_{i}(t)$
as
\begin{equation}
R^{2}(t)=\left\langle \sum_{i=1}^{9}
\frac{1}{n}\mathrm{tr}\left(\bar{A}_{i}(t)\right)^{2}\right\rangle \ ,
\label{eq:def-r}
\end{equation}
where the symbol ``tr'' represents a trace over the $n\times n$
submatrix. We can also define the ``moment of inertia tensor''
\begin{equation}
T_{ij}(t)=\frac{1}{n}\mathrm{tr}
\left(\bar{A}_{i}(t)\bar{A}_{j}(t)\right)\ ,
\label{eq:def-tij}
\end{equation}
which is a $9\times9$ real symmetric tensor. The eigenvalues of $T_{ij}(t)$
represent the spatial extent in each of the nine directions at time $t$,
and we denote them by $\lambda_{i}(t)$ with the ordering
\beq
\lambda_{1}(t)>\lambda_{2}(t)>\cdots>\lambda_{9}(t) \ .
\label{lambda-def}
\eeq
Note that $R^2(t)$ and $\lambda_i(t)$ are related as
\beq
R^2(t) = \langle \tr T \rangle =
\sum_{i=1}^9
\left\langle \lambda_i (t) \right\rangle  \ .
\label{R2-lambda-rel}
\eeq
The expectation values $\langle\lambda_{i}(t)\rangle$ can be used
as the order parameters for the spontaneous breaking of the rotational
SO(9) symmetry of the model. If the nine eigenvalues do not approach
a common value in the large-$N$ limit, we conclude 
that the SO(9) symmetry is spontaneously broken.
%, whereas some of the nine eigenvalues become larger than the others
%if the SO(9) symmetry is broken. 
From the Monte Carlo
simulations 
%with the partition function (\ref{eq:partition-function2}),
of the model (\ref{eq:partition-function2}),
it was found \cite{Kim:2011cr} that 
the three eigenvalues $\langle\lambda_{i}(t)\rangle$ ($i=1,2,3$)
start to grow with $t$
%become significantly larger than the others 
after a critical time
$t_{\mathrm{c}}$, which implies that the SO(9) symmetry is spontaneously
broken down to SO(3) for $t > t_{\rm c}$.
% in the model (\ref{eq:partition-function2}). 
%Moreover, the emergent 3d space turned out to expand with time $t$.
(See refs.~\cite{Ito:2013ywa,Ito:2015mxa}
for a precise definition of the critical time $t_{\rm c}$, 
which we use in this work.)

\subsection{Expanding behaviors in the simplified models}

%% The rate of expansion
%% is obviously an interesting quantity
%% to look at in this model.
It is interesting to investigate how the 3d space expands with time.
For that, one clearly needs to increase the matrix size, which 
is very time-consuming due to the existence of 
the Pfaffian in 
%the partition function 
(\ref{eq:partition-function2}).
%In refs.~\cite{Ito:2013ywa,Ito:2015mxa},
This led to the proposal of the simplified models,
the VDM model \cite{Ito:2013ywa}
and the bosonic model \cite{Ito:2015mxa},
which amounts to replacing the Pfaffian as
\begin{equation}
\mathrm{Pf}\mathcal{M}(A)
\Longrightarrow
\begin{cases}
\Delta(\alpha)^{16} & \text{for the VDM model} \ , \\
1 & \text{for the bosonic model} \ ,
\end{cases}\label{eq:def-vdm-bosonic}
\end{equation}
where $\Delta(\alpha)\equiv\prod_{a>b}^{N}(\alpha_{a}-\alpha_{b})$
is the van der Monde (VDM) determinant.
This replacement reduces the computational cost from $O(N^{5})$ to
$O(N^{3})$, which enables simulations with considerably large 
matrix size.
%The VDM model is expected to describe the behavior of the original model
%at early times since $\Delta(\alpha)^{16}$ 
These two models are expected to describe the qualitative behaviors 
of the original model at early times and at late times, respectively.

In both these models, the spontaneous breaking of 
the SO(9) rotational symmetry to SO(3) was observed after some
critical time as in the original model,
and the rate of expansion at late times was investigated.
% in these models.
In the VDM model, 
the extent of space $R(t)$ defined in (\ref{eq:def-r}) 
exhibits an exponential growth \cite{Ito:2013ywa}
\begin{equation}
R(t)\sim \ee^{\Lambda t}
 \ ,
%\qquad\mathrm{for}\ t>t_{\mathrm{c}} \ ,
\label{exponential}
\end{equation}
which is reminiscent of inflation\footnote{This 
behavior was observed
also in the original model \cite{Ito:2013qga} although 
the matrix size used was not large enough to 
confirm the long-time behavior.} ,
and this behavior does not seem to change with increasing $t$.
In the bosonic model, on the other hand, 
the exponential expansion observed at early times
changes into a power-law expansion \cite{Ito:2015mxa}
\begin{equation}
R(t)\sim t^{1/2} 
\label{power-law}
\end{equation}
at later times,
which is reminiscent of the Friedmann-Robertson-Walker Universe at
the radiation dominated era.
%% The rate of expansion observed in 
%% (\ref{exponential}) and (\ref{power-law}) are reminiscent of that of 
Based on these results, it has been speculated that
the extent of space $R(t)$
in the original model shows
an exponential growth at early times and
a power-law expansion at later times.
If true, it implies
that the e-folding or the duration of the cosmic inflation
may be determined dynamically in the original model.

\section{Space-time structure of the matrix configurations}
\label{sec:space-time-structure}

In this section, we investigate the space-time structure
of the matrix configurations generated by the Monte Carlo simulation
of the model 
(\ref{eq:partition-function2}) and 
the simplified models (\ref{eq:def-vdm-bosonic}).
%\subsection{how we probe the space-time structure}

\subsection{Results for the bosonic model}
\label{sec:bosonic}

%%%%%%%%%%%%%%%%%%%%%%%%%%%%
\begin{figure}
\centering{}
\includegraphics[scale=0.6]{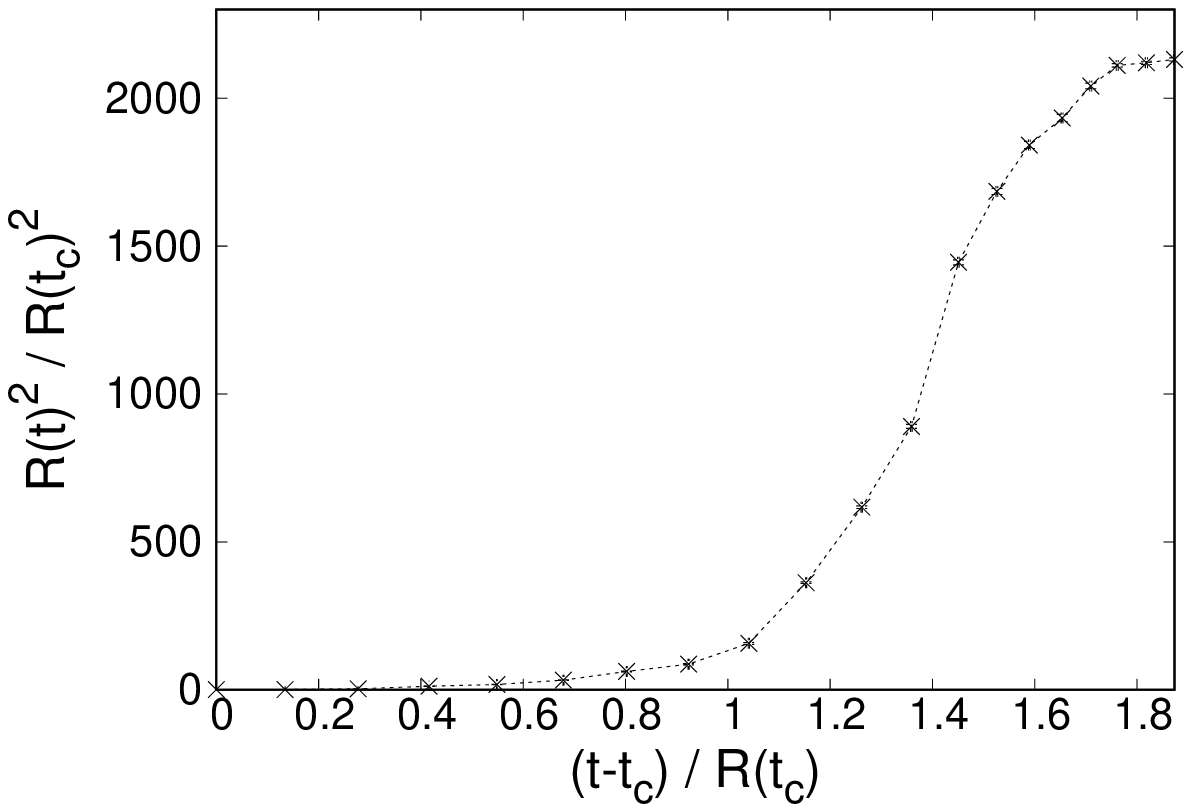}
\includegraphics[scale=0.6]{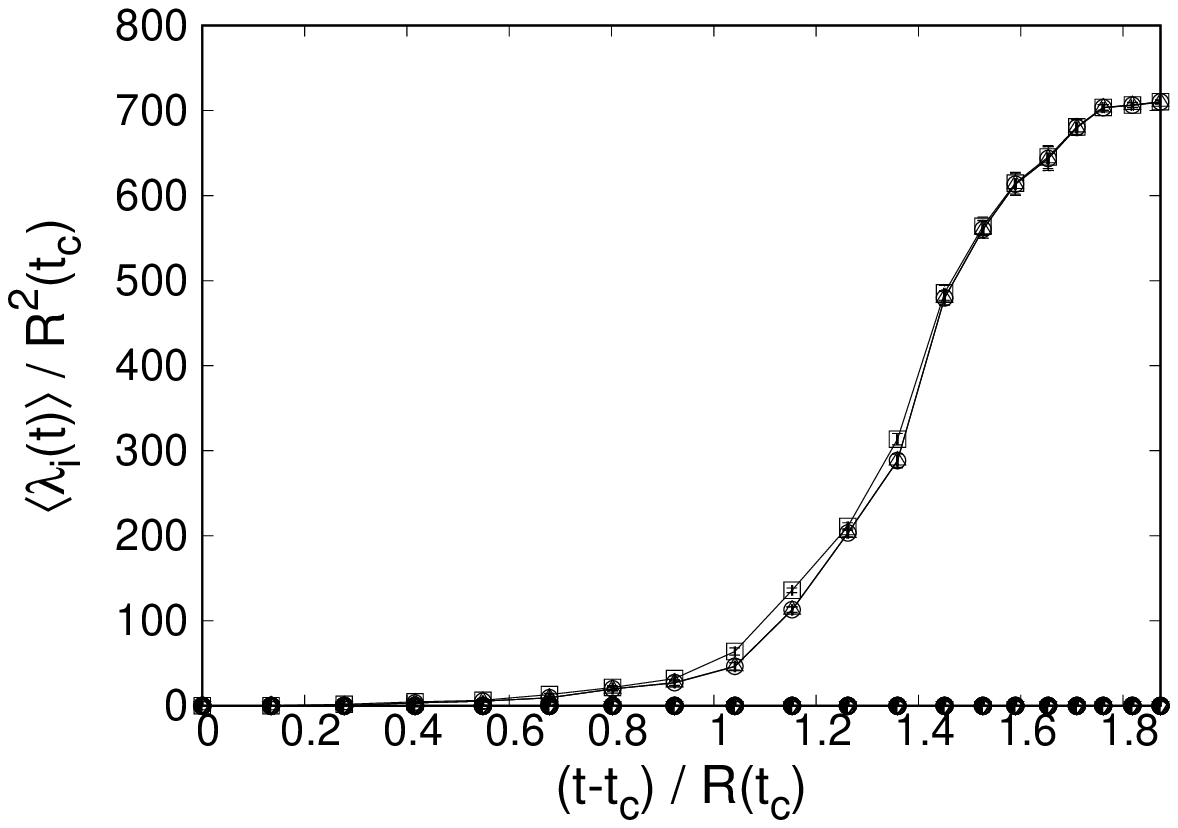}
\includegraphics[scale=0.6]{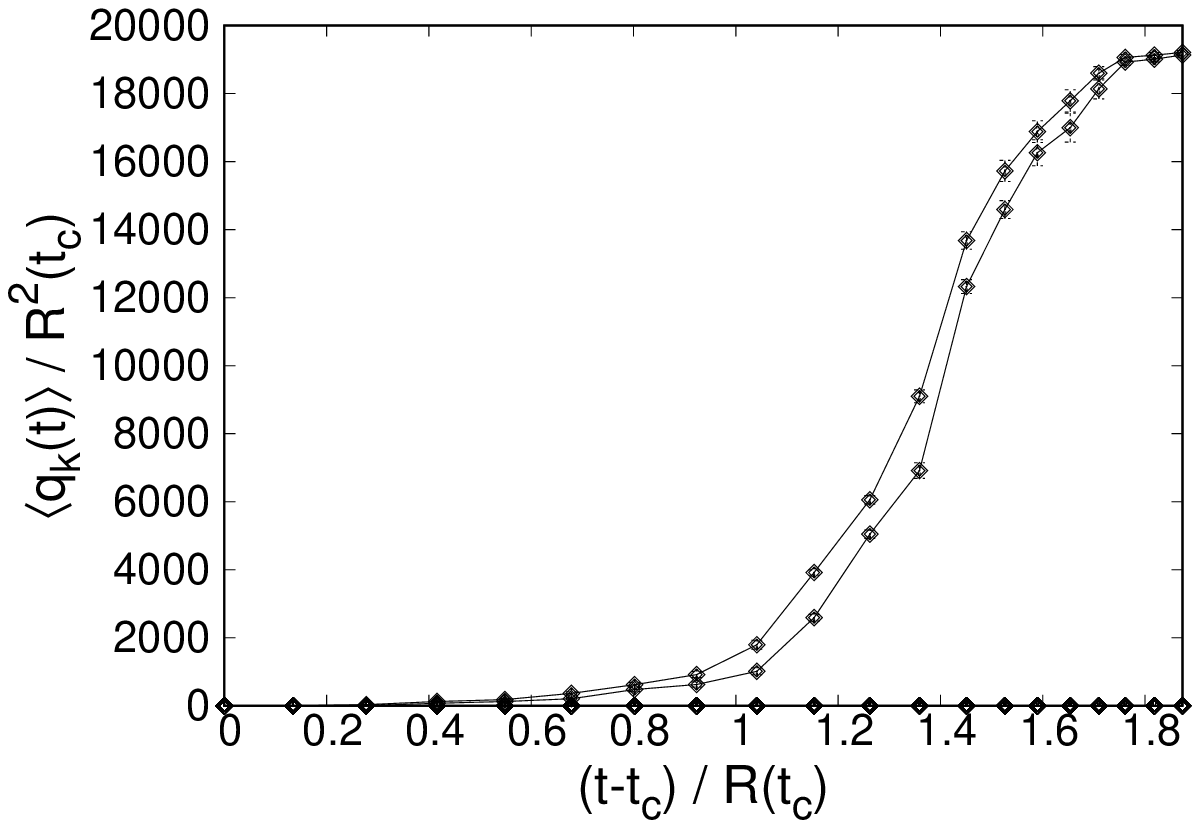}
\includegraphics[scale=0.6]{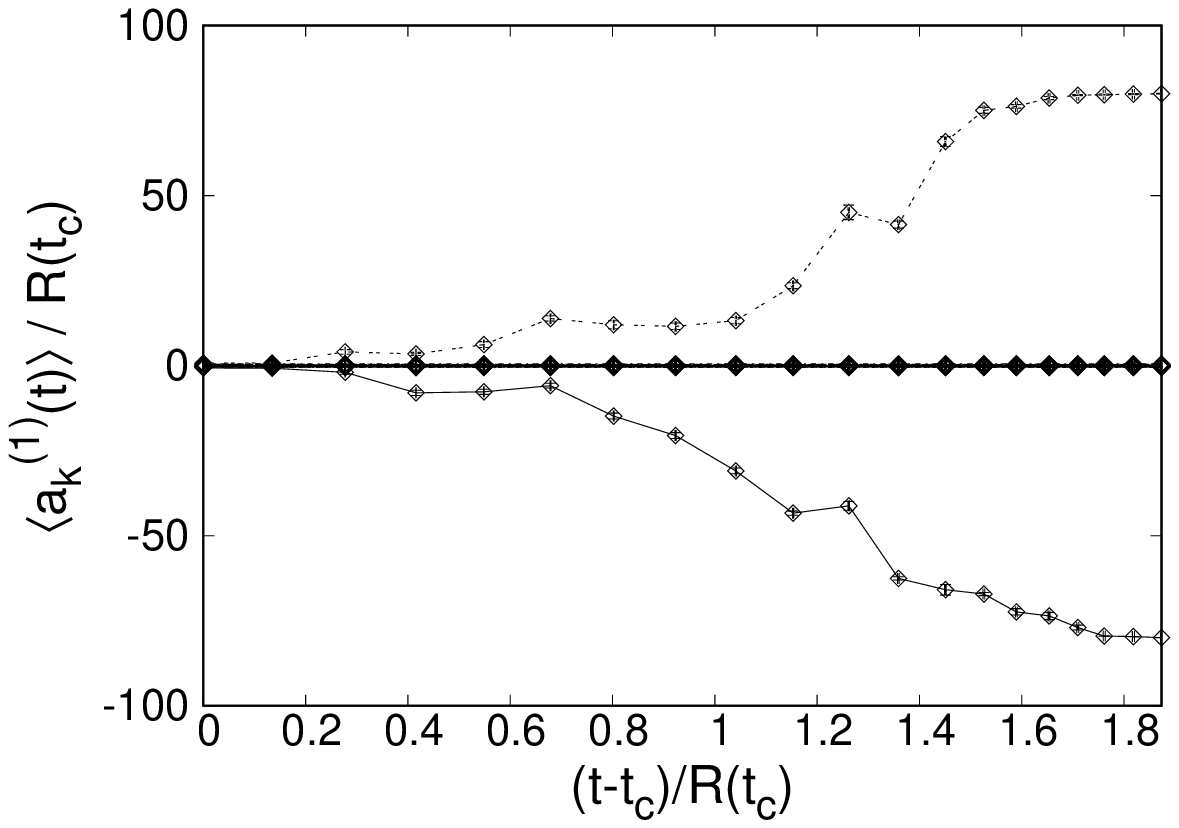}
\includegraphics[scale=0.6]{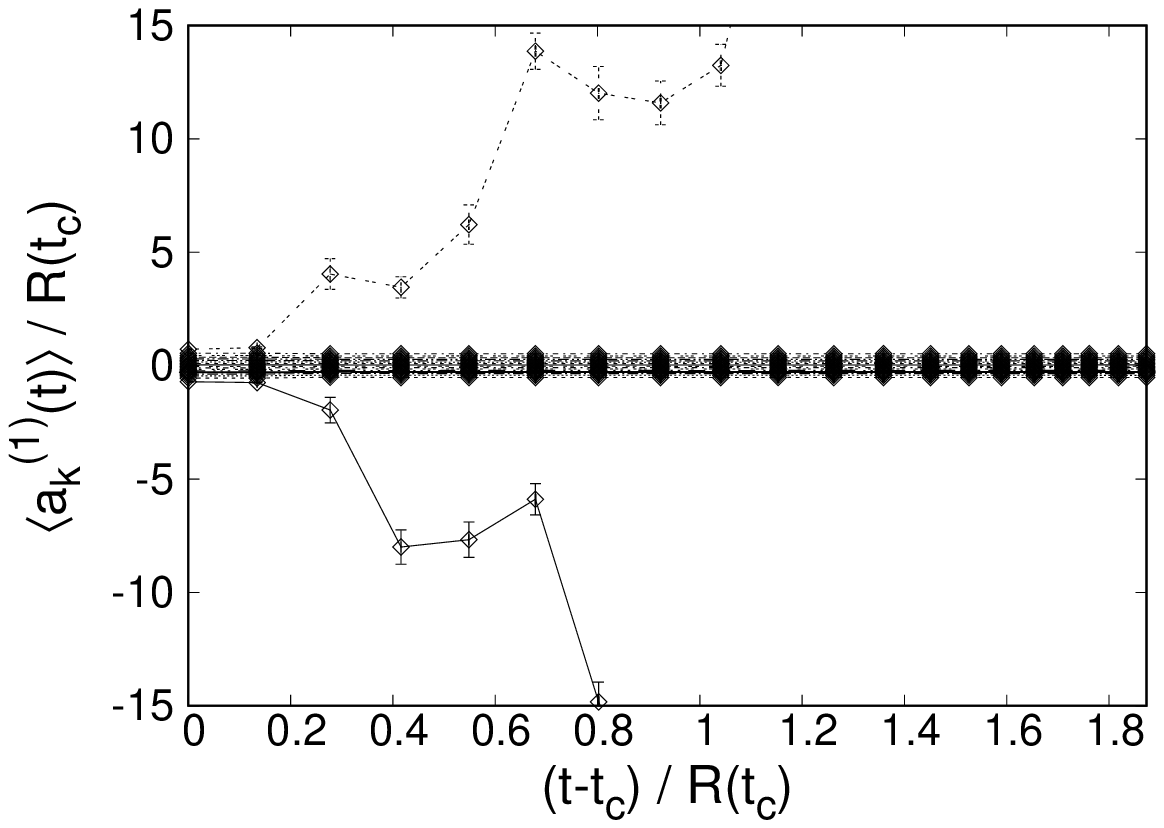}
\includegraphics[scale=0.6]{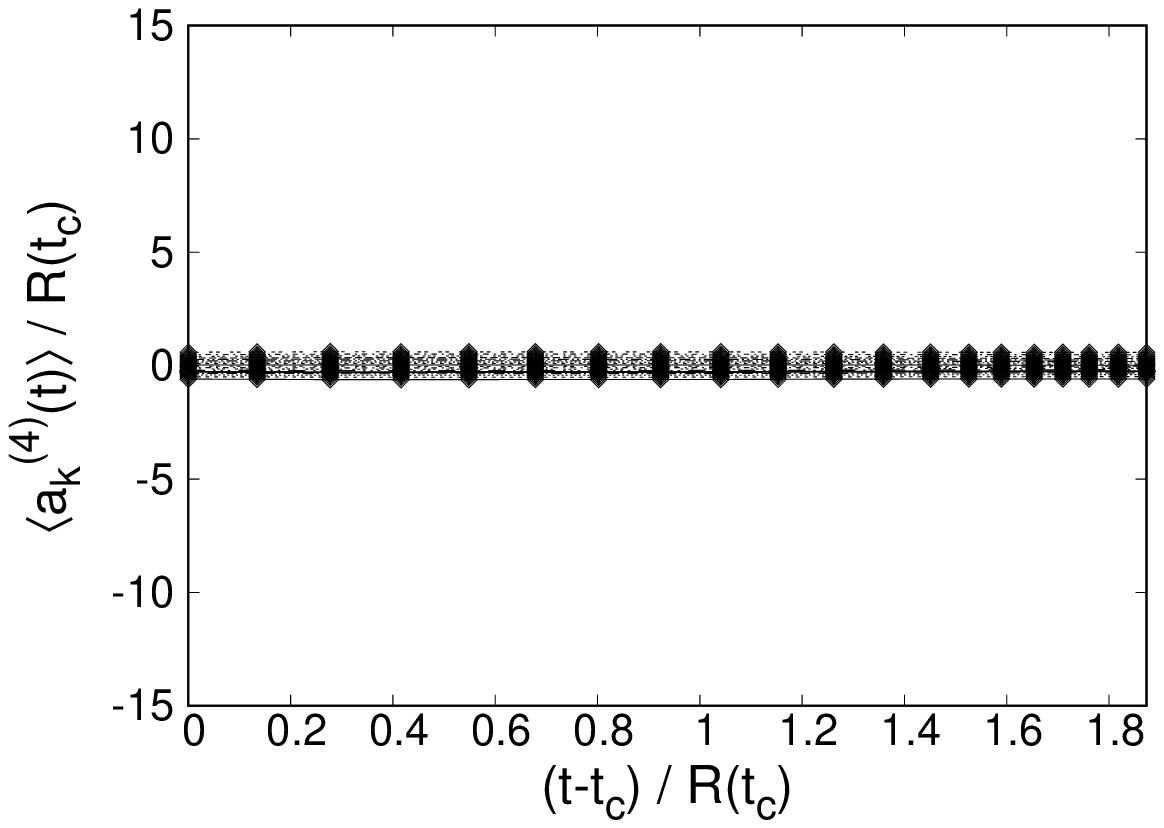}
\caption{The extent of space $R^{2}(t)/R^{2}(t_{\rm c})$ (Top-Left) 
and 
the normalized eigenvalues $\langle \lambda_i(t) \rangle /R^{2}(t_{\rm c})$
of $T_{ij}(t)$ (Top-Right)
are plotted
against time $(t-t_{\rm c})/R(t_{\rm c})$
for the bosonic model with $N=256$, $C=100$, $\kappa=1$, $p=1.5$
and the block size $n=18$.
Similarly, 
the eigenvalues of $Q(t)/R^{2}(t_{\mathrm{c}})$ 
(Middle-Left),
the eigenvalues of $\bar{A}^{(1)}(t)/R(t_{\rm c})$
(Middle-Right, Bottom-Left, the latter being the zoom-up version 
of the former),
the eigenvalues of $\bar{A}^{(4)}(t)/R(t_{\rm c})$
(Bottom-Right)
are plotted against time $(t-t_{\rm c})/R(t_{\rm c})$.}
\label{fig:R2_bosonic|eigen_a_bosonic} 
\end{figure}
%%%%%%%%%%%%%%%%%%%%%%%%%%%%%%

In this subsection, we consider the bosonic model, 
which is a simplified model for the late time behaviors.
Let us first look at the basic quantities such as
the extent of space $R^{2}(t)$
and the eigenvalues $\langle \lambda_i (t) \rangle$ of $T_{ij}(t)$.
In Fig.~\ref{fig:R2_bosonic|eigen_a_bosonic}
%\ref{fig:R2_bosonic|eigen_a_bosonic}(left),
we plot the extent of space $R^{2}(t)/R^{2}(t_{\mathrm{c}})$ (Top-Left)
and 
the normalized eigenvalues 
$\langle \lambda_i (t) \rangle/R^{2}(t_{\mathrm{c}})$ of $T_{ij}(t)$ 
(Top-Right) 
%defined in (\ref{eq:def-tij})
against
$(t-t_{\mathrm{c}})/R(t_{\mathrm{c}})$ for 
$N=256$, $C=100$, $\kappa=1.0$ with the block size $n=18$
in (\ref{eq:def-r}).
Here and for all the other plots in 
Fig.~\ref{fig:R2_bosonic|eigen_a_bosonic},
we only present the results in the $t<0$ region 
since the results are symmetric\footnote{This does not
mean that the Big Crunch occurs in this model
because the time difference between the symmetric point $t=0$ 
and the critical time $t=t_{\rm c}$ seems to diverge
in physical units in an appropriate large-$N$ limit.
See section \ref{sec:cont-lim}.} under the 
time reflection $t \mapsto - t$.
%% In order to probe the late time behaviors efficiently,
%% we tune the parameter $C$ in
%% (\ref{eq:partition-function2})
%% appropriately following ref.~\cite{Azuma:2017dcb}.
%
%The normalization of each quantity is chosen in such a way
%that 
The power $p$ in the IR cutoff (\ref{eq:cutoff-temp}) and 
(\ref{eq:cutoff-space}) is chosen to be $p=1.5$, which
is found to be large enough to make 
the results almost independent of $p$ (See 
Appendix \ref{sec:appendix_universality}.).
Let us recall that 
$R^{2}(t)$
is related to $\langle \lambda_i (t) \rangle$
through (\ref{R2-lambda-rel}).
While the extent of space $R^{2}(t)/R^{2}(t_{\mathrm{c}})$
grows with $t$ for $t > t_{\rm c}$,
it is only three out of nine eigenvalues of $T_{ij}(t)$
that grow with $t$, which suggests that the rotational SO(9) symmetry
is broken spontaneously to SO(3).
These results are analogous to the previous results 
obtained for $p=1$ \cite{Ito:2015mxa}.

% for $t > t_{\rm c}$.
%% Below we focus on the behavior obtained after the critical 
%% time $t_{\rm c}$, at which the SO(9) symmetry is
%% spontaneously broken.

%% In Fig.~\ref{fig:R2_bosonic|eigen_a_bosonic} (Top-Right),
%% we plot the eigenvalues $\langle \lambda_i (t) \rangle$ of $T_{ij}(t)$
%% defined in (\ref{eq:def-tij}).
%% % against $(t-t_{\mathrm{c}})/R(t_{\mathrm{c}})$.

The simplest way to probe the space-time structure
is to define an $n\times n$ matrix
\begin{equation}
Q(t)\equiv\sum_{i=1}^{9}\ (\bar{A}_{i}(t))^{2} ,
\label{eq.2}
\end{equation}
which is invariant under SO(9) rotations.
Let us denote its eigenvalues as $q_{k}(t)$ ($k=1, \cdots , n$) 
with the ordering 
\beq
q_1(t) < \cdots < q_n(t) \ .
\label{q-ordering-def}
\eeq
%$q_1(t) < \cdots < q_n(t)$.
%% The eigenvalues $q_{k}(t)$ ($k=1, \cdots , n$) of 
%% this matrix $Q(t)$
%$Q(t)$
These eigenvalues
tell us how the space spreads in the radial direction at each time $t$.

In Fig.~\ref{fig:R2_bosonic|eigen_a_bosonic} (Middle-Left),
we plot the eigenvalues $q_{k}(t)/R^2(t_{\mathrm{c}})$ 
against $(t-t_{\rm c})/R(t_{\rm c})$.
We find that the two largest eigenvalues grow with $t$, but not the others.
Let us note that the eigenvalues of $Q(t)$
% and those of $\bar{A}^{(i)}(t)$
are related to the extent of space $R^2(t)$ as
\beq
R^2(t) = 
\left\langle \frac{1}{n}\tr Q(t) \right\rangle 
= \left \langle \frac{1}{n} \sum_{k=1}^n  q_k(t) \right \rangle \ .
%= \left \langle  \sum_{i=1}^{9} \sum_{k=1}^n  \Big( a^{(i)}_{k}(t) \Big)^2 
% \right \rangle \  .
\label{R2-q-eigen-rel}
\eeq
This implies that
%the eigenvalues except for $q_{k}(t)$
the time-dependence of $R^2(t)$ 
seen in the Top-Left panel
is caused only by
the two largest eigenvalues of $Q(t)$.

Let us next discuss the space-time structure
in the three extended directions and the six shrunken directions 
separately.
Since we are dealing with spontaneous symmetry breaking,
we need to choose the frame properly
in order to distinguish these directions.
%the three extended directions and the six shrunken directions.
% to distinguish
%the extended directions and the shrunken directions.
Suppose $v_j^{(i)}(t)$ ($j=1,\cdots ,9$) are the normalized 
eigenvectors of 
the ``moment of inertia tensor'' (\ref{eq:def-tij}) 
corresponding to the eigenvalues $\lambda_i(t)$ 
with the ordering (\ref{lambda-def}).
Then, we can define the 
$n \times n$ matrix
corresponding to the 
spatial direction with the extent $\lambda_i$ as
\beq
\bar{A}^{(i)}(t) =  \sum_{j=1}^{9}  v_j^{(i)}(t) \,  \bar{A}_j(t)
\eeq
and its eigenvalues
%$\bar{A}^{(i)}(t)$ by 
$a^{(i)}_k(t)$ ($k=1, \cdots ,n$) with the ordering 
\begin{equation}
a_{1}^{(i)}(t)<\cdots< a_{n}^{(i)}(t)\ .
\label{eq:EigenvaluesOfA}
\end{equation}

In Fig.~\ref{fig:R2_bosonic|eigen_a_bosonic} (Middle-Right),
we plot the eigenvalues $a^{(1)}_{k}(t)/R(t_{\mathrm{c}})$ against
$(t-t_{\mathrm{c}})/R(t_{\mathrm{c}})$.
We find that only
two eigenvalues $a_{1}^{(1)}(t)$ and $a_{n}^{(1)}(t)$ 
grow in magnitude with time $t$,
and all the others remain close to zero.
Similar behaviors are seen also for the eigenvalues
$a^{(2)}_{k}(t)$ and $a^{(3)}_{k}(t)$ obtained 
for the other 
extended directions.
In Fig.~\ref{fig:R2_bosonic|eigen_a_bosonic} (Bottom-Left),
we zoom up the same plot to make visible 
the eigenvalues close to zero. 
In Fig.~\ref{fig:R2_bosonic|eigen_a_bosonic} (Bottom-Right),
we plot the eigenvalues $a^{(4)}_{k}(t)/R(t_{\mathrm{c}})$ against
$(t-t_{\mathrm{c}})/R(t_{\mathrm{c}})$.
We find that all the eigenvalues
remain close to zero.
Similar behaviors are seen also for the eigenvalues
$a^{(5)}_{k}(t), \cdots , a^{(9)}_{k}(t)$ obtained for 
the other shrunken directions.
Comparing the two plots at the bottom of 
Fig.~\ref{fig:R2_bosonic|eigen_a_bosonic},
%the results for the extended directions with those for
%the shrunken directions,
we notice that the eigenvalue distribution of $\bar{A}^{(i)}$
is almost identical 
for the extended directions and the shrunken directions
except for the two eigenvalues with large magnitude.

Similarly to (\ref{R2-q-eigen-rel}), 
the eigenvalues of $\bar{A}^{(i)}(t)$
%are related to the extent of space $R^2(t)$ as
are related to the extent of space $\lambda_i(t)$ in the $i$th direction as
\beq
\lambda_i (t)
%\langle \frac{1}{n}\tr Q(t) \rangle 
%\left \langle \sum_{k=1}^n \Big( q_k(t) \Big)^2 \right \rangle
= \frac{1}{n} \sum_{k=1}^n  \Big( a^{(i)}_{k}(t) \Big)^2  \ .
\label{lambda-a-eigen-rel}
\eeq
Our observation implies that
%the eigenvalues except for $q_{k}(t)$
%% the time-dependence of $\lambda_i(t)$ for the extended directions $i=1,2,3$
%% is caused only by the two eigenvalues of $\bar{A}^{(i)}(t)$ with the large
%% magnitude.
the spontaneous symmetry breaking of the SO(9) 
rotational symmetry 
seen in the Top-Right panel
is caused only
by the two eigenvalues of $\bar{A}^{(i)}(t)$ with large
magnitude.
%Thus we find that the space-time structure of the matrix configurations
%obtained in the simulation is actually singular.

%%%%%%%%%%%%%%%%%%%%%%%%%%%

\subsection{Including fermionic contributions}
\label{sec:fermions}

%%%%%%%%%%%%%%%%%%%%%%%%%%%%%%%%%%%

In order to seek for the possibility to obtain a regular space-time,
%we investigate the space-time structure
we repeat the analysis in the previous subsection
in the case of the original model (\ref{eq:partition-function2})
including fermionic contributions.
Since the cost of Monte Carlo simulations increases
from O($N^3$) to O($N^5$),
% from the bosonic model,
% discussed in the previous section, 
here we restrict ourselves to a rather small matrix size $N=16$.

\begin{figure}
\centering{}
\includegraphics[scale=0.6]{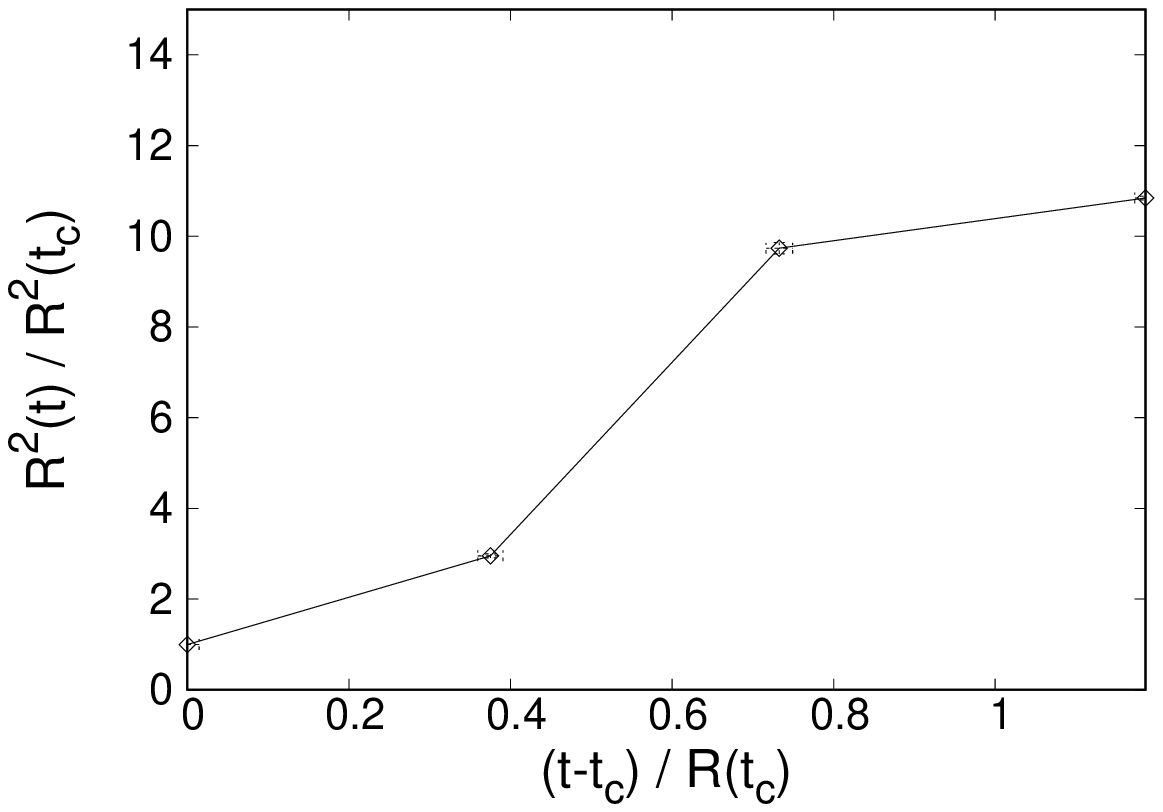}
\includegraphics[scale=0.6]{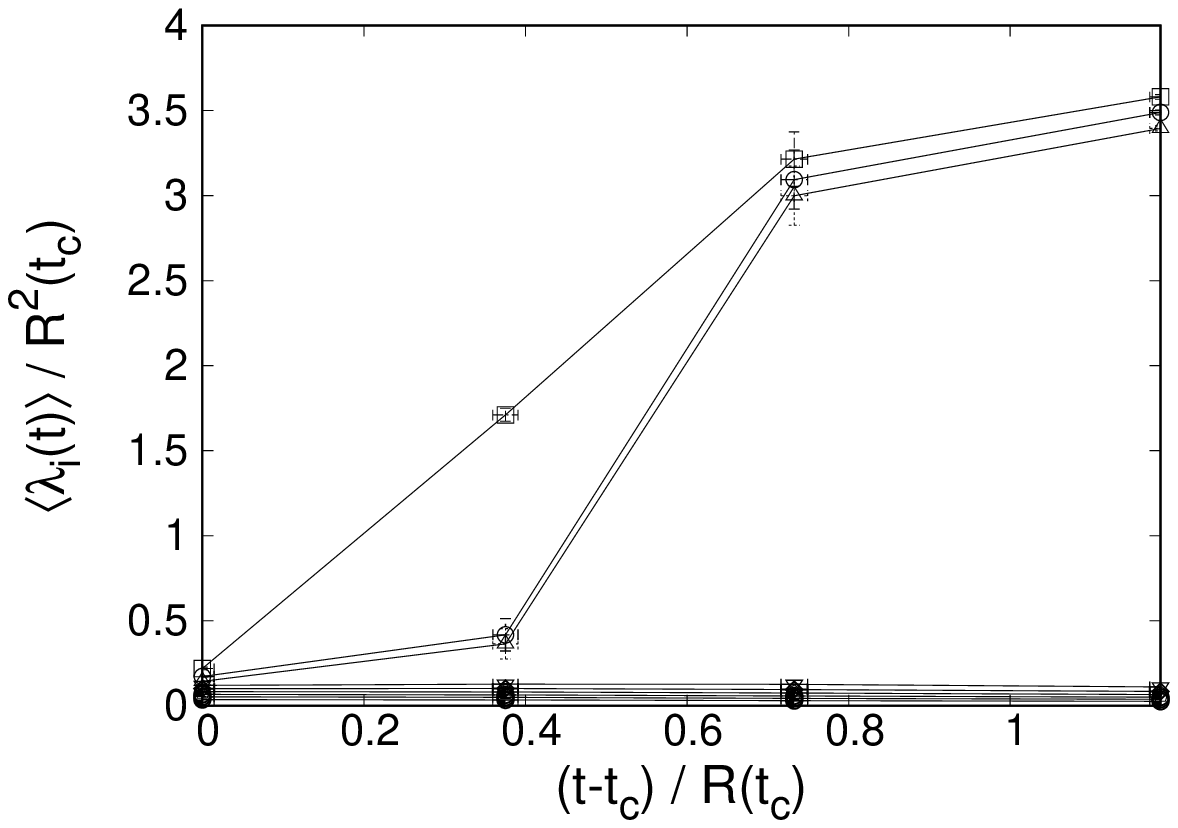}
\includegraphics[scale=0.6]{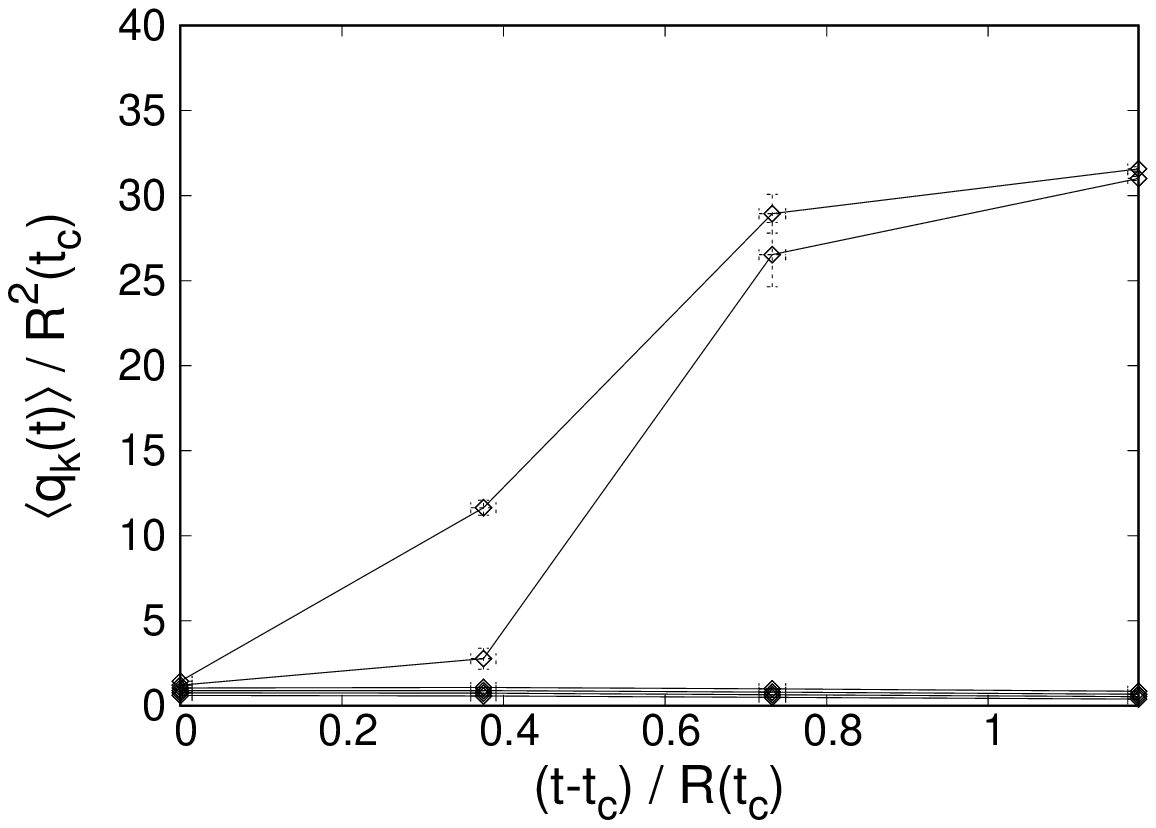}
\includegraphics[scale=0.6]{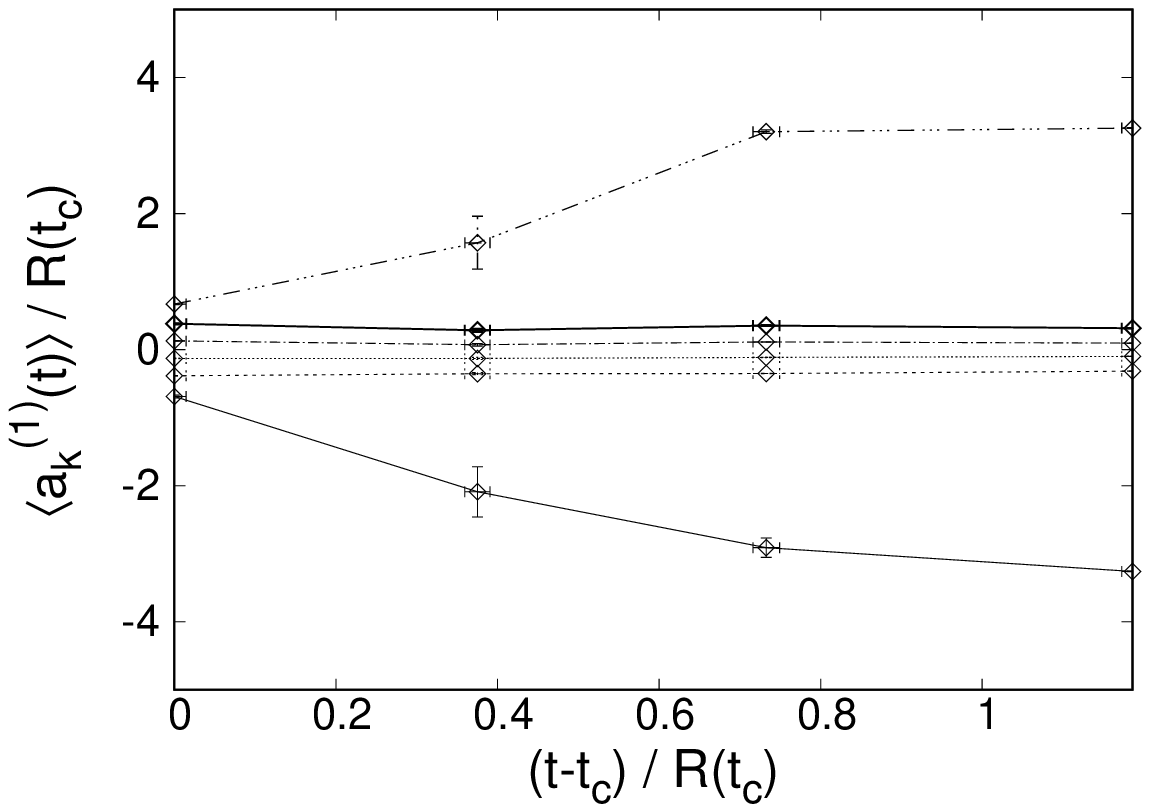}
\includegraphics[scale=0.6]{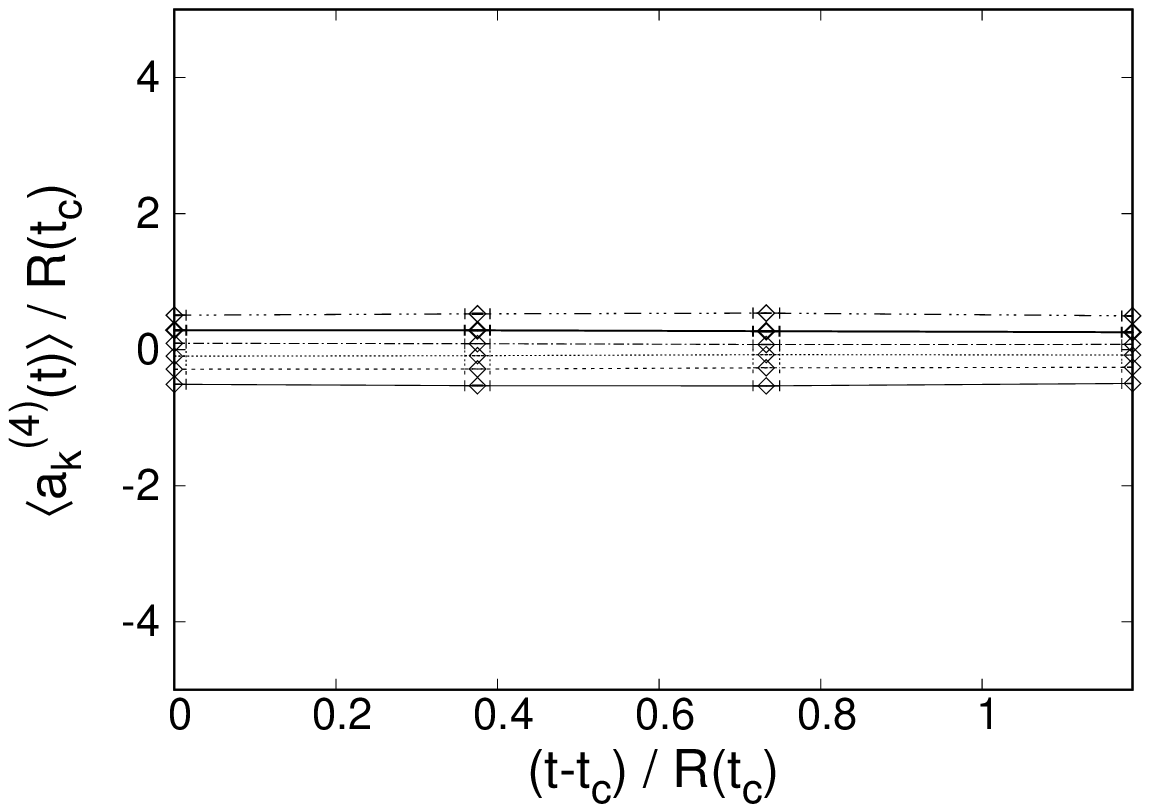}
\caption{The extent of space $R^{2}(t)/R^{2}(t_{\rm c})$ (Top-Left)
and
the normalized 
eigenvalues $\langle \lambda_i(t) \rangle /R^{2}(t_{\rm c})$
of $T_{ij}(t)$ (Top-Right)
are plotted
against time $(t-t_{\rm c})/R(t_{\rm c})$
for the original model with $N=16$, $C=3.91$, $\kappa=0.38$, $p=1.6$
and the block size $n=6$.
Similarly,
the eigenvalues of $Q(t)/R^{2}(t_{\mathrm{c}})$ (Middle-Left),
the eigenvalues of $A^{(1)}(t)/R(t_{\rm c})$ (Middle-Right) and
the eigenvalues of $A^{(4)}(t)/R(t_{\rm c})$ (Bottom)
are plotted against time $(t-t_{\rm c})/R(t_{\rm c})$.
}
\label{fig:R2_ikkt|eigen_a_ikkt} 
\end{figure}

%% We tune the parameter $C$ 
%% in (\ref{eq:partition-function2})
%% appropriately following ref.~\cite{Azuma:2017dcb}.
In Fig.~\ref{fig:R2_ikkt|eigen_a_ikkt}
we plot the same quantities 
as in Fig.~\ref{fig:R2_bosonic|eigen_a_bosonic} 
for the original model
with $N=16$, $C=3.91$, $\kappa=0.38$ and the block size $n=6$. 
The power $p$ in the IR cutoff (\ref{eq:cutoff-temp}) and 
(\ref{eq:cutoff-space}) is chosen to be $p=1.6$, which
is found to be large enough
to make the results almost independent 
of $p$ (See Appendix \ref{sec:appendix_universality}.).
These results 
are qualitatively
the same as those obtained for the bosonic model. 
While the fermionic matrices are expected to play an important role
in the properties of the model such as the expanding behavior, 
they do not seem to affect the singular space-time structure.
%% (Top-Left), we plot the extent
%% of space $R^{2}(t)/R^{2}(t_{\mathrm{c}})$ against 
%% time $(t-t_{\mathrm{c}})/R(t_{\mathrm{c}})$
%% for the original model
%% with $N=16$, $C=3.91$ and $\kappa=0.38$. 
%% The power $p$ in the IR cutoff (\ref{eq:cutoff-temp}) and 
%% (\ref{eq:cutoff-space}) were chosen to be $p=1.6$, at which
%% the results are found to be almost independent of $p$.
%% (See Appendix \ref{sec:appendix_universality}.)

%% In Fig.~\ref{fig:R2_ikkt|eigen_a_ikkt} 
%% we also plot
%% the eigenvalues $q_{k}(t)$ ($k=1, \cdots , n$)
%% of the matrix $Q(t)$ (Top-Right),
%% the eigenvalues of $\bar{A}_{1}(t)$ (Bottom-Left)
%% and the eigenvalues of $\bar{A}_{4}(t)$ (Bottom-Right)
%% against time $(t-t_{\rm c})/R(t_{\rm c})$.

%% \begin{figure}
%% \centering{}
%% \includegraphics[scale=0.6]{IKKT_bulk}
%% \includegraphics[scale=0.6]{IKKT_Q}
%% \caption{ These plots are obtained by using the SUSY model with $N=16,C=3.91$,
%% and $\kappa=0.38$. (Left) $\lambda_{i}^{\mathrm{bulk}}(t)/R(t_{\mathrm{c}})^{2}$
%% are plotted against time $(t-t_{\mathrm{c}})/R(t_{\mathrm{c}})$.
%% In (Right), the eigenvalues $q_{k}(t)/R(t_{\mathrm{c}})^{2}$ for
%% various time $t$ are plotted against $k/(n-1)$. The order of the
%% time $t$ is $t_{1}<t_{2}<t_{3}<t_{4}$ with $t_{0}=t_{\mathrm{c}}$
%% and $t_{4}=t_{\mathrm{peak}}$ . }
%% \label{fig:lam_bulk_and_Q_eigen_IKKT} 
%% \end{figure}

%%%%%%%%%%%%%%%%%%%%%%%%%%%%%%%%%%%
\subsection{Taking the continuum limit}
\label{sec:cont-lim}
%Space-time structure in the continuum limit

\begin{figure}
\centering{}
%\begin{centering}
\includegraphics[scale=0.6]{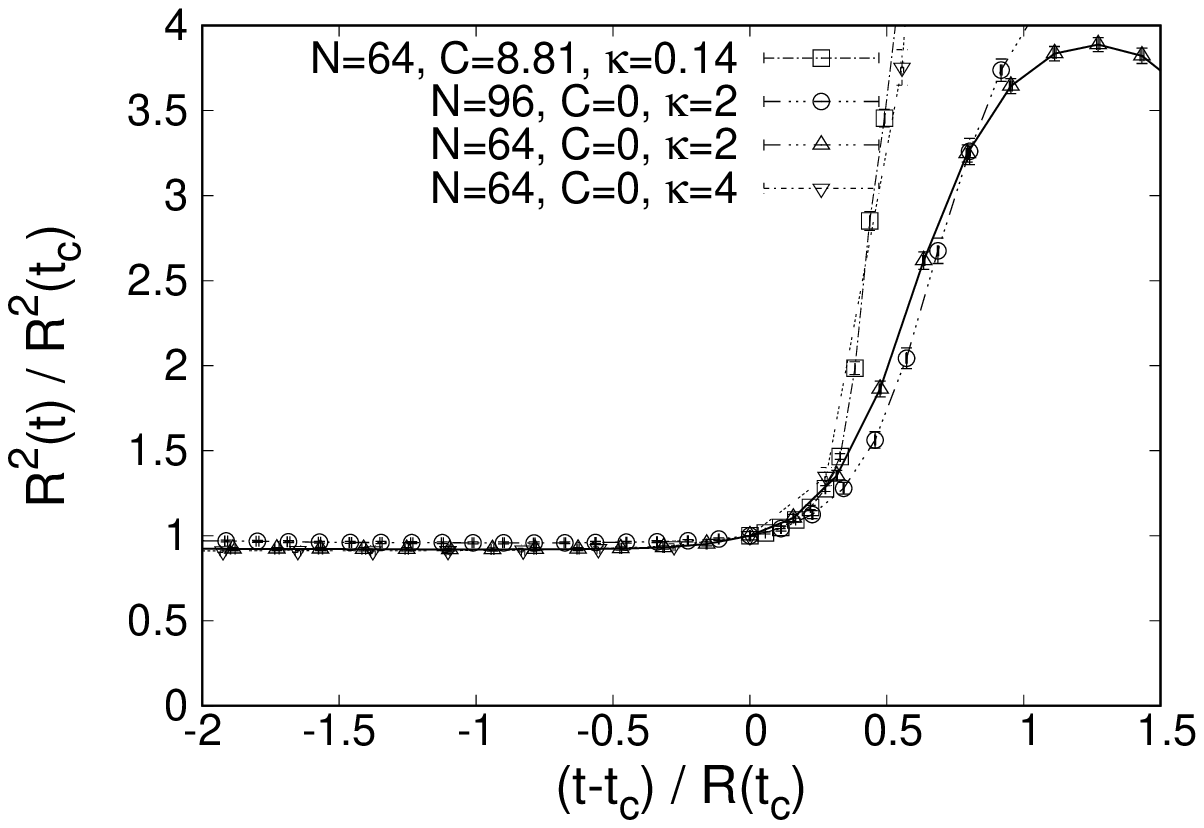}
\includegraphics[scale=0.6]{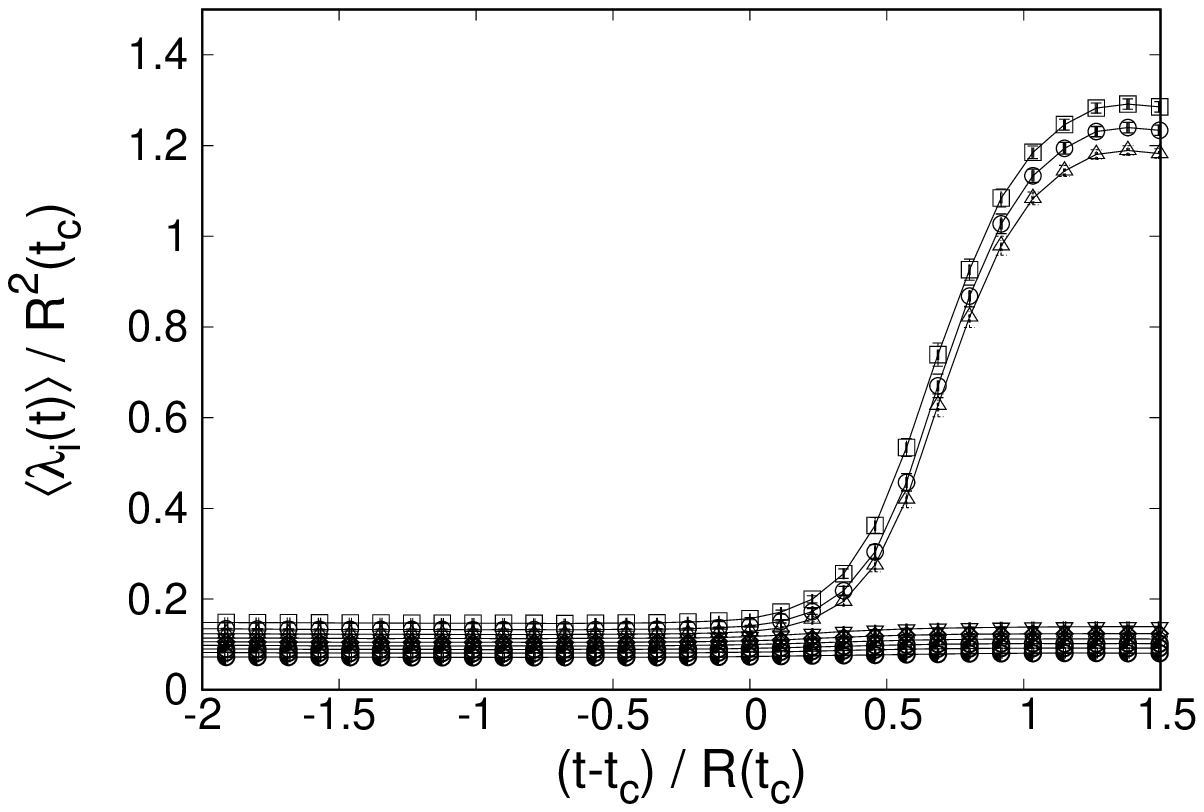}
\includegraphics[scale=0.6]{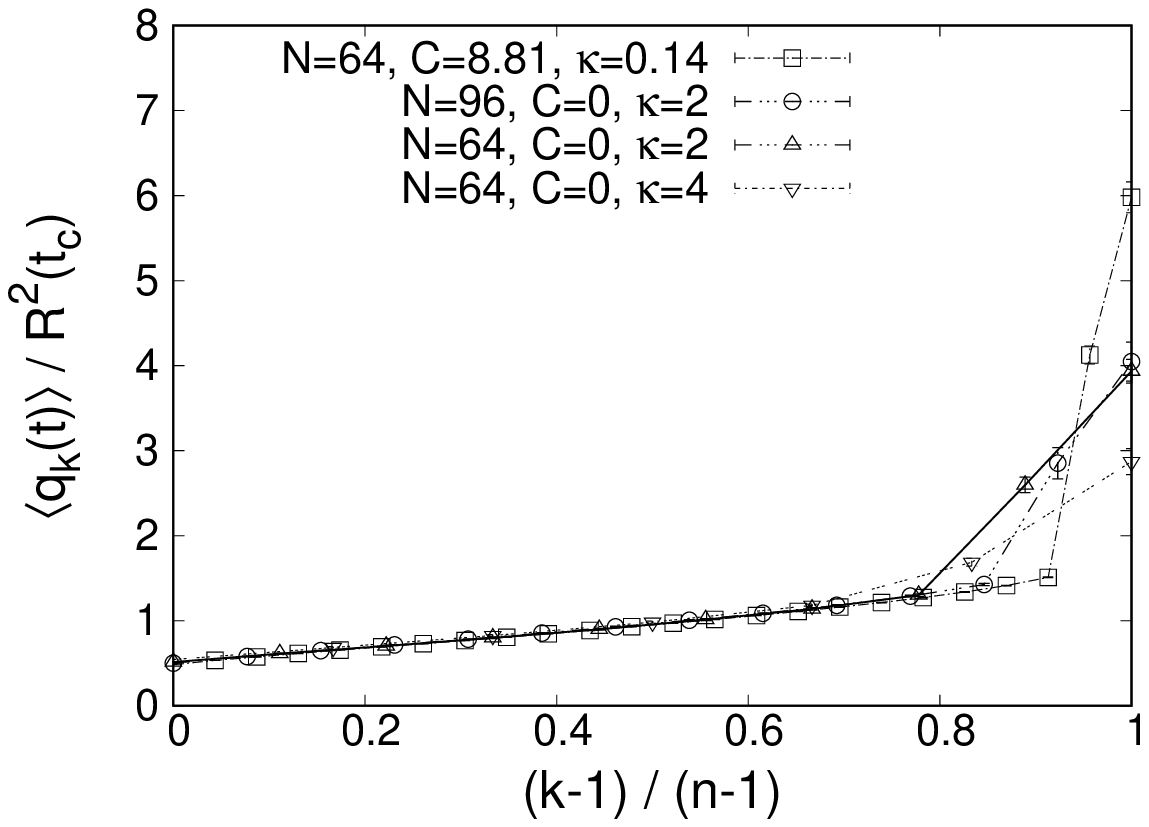}
%\par\end{centering}
%\centering{}
\includegraphics[scale=0.6]{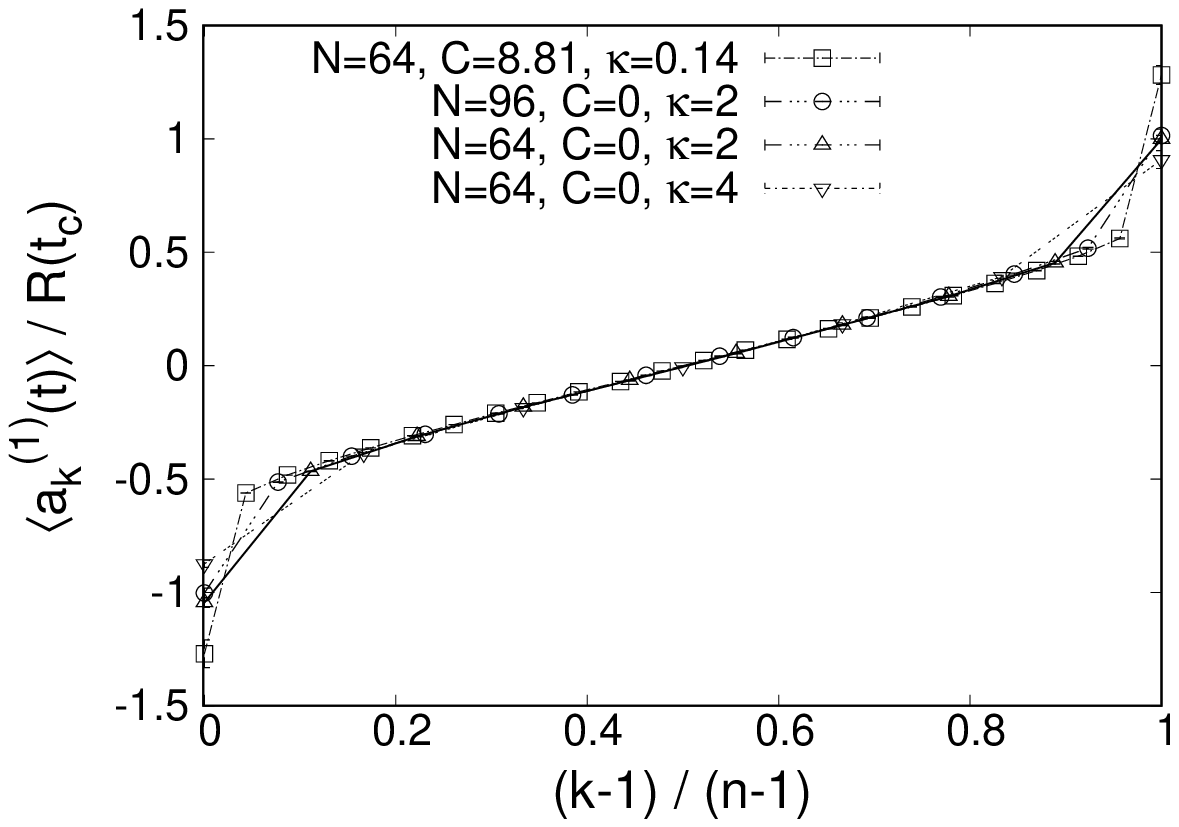}
\includegraphics[scale=0.6]{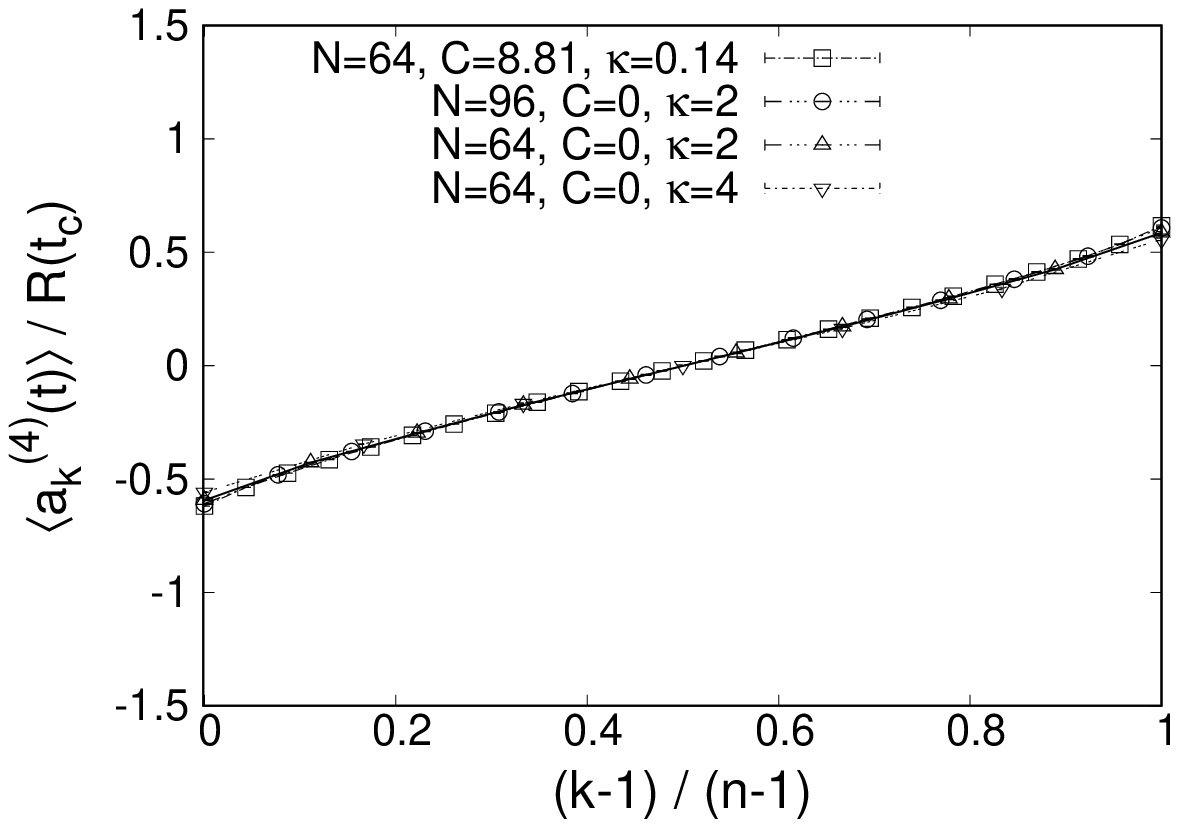}
\caption{(Top-Left) The extent of space $R^{2}(t)/R^{2}(t_{\rm c})$
is plotted against time $(t-t_{\rm c})/R(t_{\rm c})$ 
for the VDM model with the parameter sets ($N$, $C$, $\kappa$)
and the block size $n$ listed in table \ref{InformationInVDM}.
The power $p$ in the IR cutoff (\ref{eq:cutoff-temp}) and 
(\ref{eq:cutoff-space}) is chosen as $p=1.4$.
%% The ``volume'' $\Delta$  and the ``lattice spacing'' $\epsilon$ 
%% obtained for each parameter set,
%% as well as the adopted block size $n$,
%% are given in table~\ref{InformationInVDM}. 
(Top-Right)
The normalized
eigenvalues $\langle \lambda_i(t) \rangle /R^{2}(t_{\rm c})$
of $T_{ij}(t)$
are plotted
against time $(t-t_{\rm c})/R(t_{\rm c})$
for $N=96$, $C=0$, $\kappa=2$.
The eigenvalues 
%$q_{k}(t)/R^{2}(t_{\mathrm{c}})$
of $Q(t)/R^{2}(t_{\mathrm{c}})$ (Middle-Left),
the eigenvalues
of $\bar{A}^{(1)}(t)/R(t_{\rm c})$ (Middle-Right)
and the eigenvalues of $\bar{A}^{(4)}(t)/R(t_{\rm c})$ (Bottom)
obtained at $(t-t_{\rm c})/R(t_{\rm c})\sim0.40$
are plotted against
their labels $(k-1)/(n-1)$
for the four parameter sets listed in table~\ref{InformationInVDM}. 
}
\label{fig:vdm10dp1point4_t_r2|eigen_a_vdm} 
\end{figure}

As yet another possibility to obtain a regular space-time,
let us consider taking the continuum limit.
Here we use the VDM model, which is a simplified model
for the early time behaviors.
In Fig.~\ref{fig:vdm10dp1point4_t_r2|eigen_a_vdm} (Top-Left), we 
plot the extent of space $R^{2}(t)/R^{2}(t_{\mathrm{c}})$ against 
time $(t-t_{\mathrm{c}})/R(t_{\mathrm{c}})$
%for the VDM model 
for various $N$, $C$ and $\kappa$ 
with the block size $n$ 
listed in table~\ref{InformationInVDM}.
The power $p$ in the IR cutoff (\ref{eq:cutoff-temp}) and 
(\ref{eq:cutoff-space}) is chosen as $p=1.4$ following
%, at which the $p$-dependence of the results 
%is found to be negligible 
ref.~\cite{Ito:2017rcr}.
From this plot,
we observe a clear scaling behavior 
for $(t-t_{\rm c})/R(t_{\rm c}) \lesssim  0.40$.

In Fig.~\ref{fig:vdm10dp1point4_t_r2|eigen_a_vdm} (Top-Right),
we plot the normalized
eigenvalues $\langle \lambda_i (t) \rangle /R^{2}(t_{\mathrm{c}})$ 
of $T_{ij}(t)$
%defined in (\ref{eq:def-tij})
for the VDM model with $N=96$, $C=0$ and $\kappa=2$.
% against $(t-t_{\mathrm{c}})/R(t_{\mathrm{c}})$.
Similar behaviors are obtained for the other parameter sets.
We find that three out of nine eigenvalues of $T_{ij}(t)$
grow with time, which suggests that the rotational SO(9) symmetry
is broken spontaneously to SO(3) for $t> t_{\rm c}$.
These results are similar to those obtained
in refs.~\cite{Ito:2013ywa,Ito:2017rcr}.

\begin{table}
\centering %
\begin{tabular}{|c|c|c||c|c|c|}
\hline 
$N$  & $C$  & $\kappa$  & $n$  & $\Delta$  & $\epsilon$\tabularnewline
\hline 
\hline 
64  & 8.81  & 0.14  & 24  & 1.0990(16)  & 0.0550(1)\tabularnewline
\hline 
96  & 0  & 2.00  & 14 & 1.3811(41)  & 0.1151(3)\tabularnewline
\hline 
64  & 0  & 2.00  & 10  & 1.2726(63)  & 0.1591(8)\tabularnewline
\hline 
64 & 0 & 4.00 & 7 & 1.3762(87) & 0.2752(17)\tabularnewline
\hline 
\end{tabular}\caption{The parameter sets 
($N$, $C$, $\kappa$) used for the simulation of the VDM model
are listed.
%of the submatrices representing each time slice are listed.
We also present
the block size $n$,
the ``volume'' $\Delta$  and the ``lattice spacing'' $\epsilon$ 
determined from the data for each parameter set.}
\label{InformationInVDM} 
\end{table}

%%%%%%%%%%%%%%%%%%%%%%%%%%%%%%

%%%%%%%%%%%%%%%%%%%%%%%%%%%%%%

In order to discuss the continuum limit, let us
define the ``volume'' and the ``lattice spacing''
in the temporal direction as \cite{Ito:2013ywa}
\beq
\Delta  \equiv
\frac{t_{\mathrm{peak}}-t_{\rm c}}{R\left(t_{\mathrm{c}}\right)} \ , 
\quad \quad
\epsilon \equiv\frac{\Delta}{\nu}\ ,
\label{eq:def_eps}
\eeq
%% \begin{align}
%% \Delta & \equiv
%% \frac{t_{\mathrm{peak}}-t_{\rm c}}{R\left(t_{\mathrm{c}}\right)} \ ,
%% \label{eq:def_delta}
%% \\
%% \epsilon & \equiv\frac{\Delta}{\nu}\ ,
%% \label{eq:def_eps}
%% \end{align}
where $t_{\mathrm{peak}}$ represents the position of the peak in
$R^{2}(t)$
and $\nu$ is the number of data points of $R^{2}(t)$ contained
within $t_{\rm c}<t\leq t_{\mathrm{peak}}$. 
Roughly speaking, the lattice spacing $\epsilon$ 
represents the average horizontal spacing between the adjacent data points of
$R^{2}(t)/R^{2}(t_{\mathrm{c}})$. 
In table~\ref{InformationInVDM}, we present
the volume $\Delta$ and 
the lattice spacing $\epsilon$
obtained for each parameter set $(N, C,\kappa)$
used in Fig.~\ref{fig:vdm10dp1point4_t_r2|eigen_a_vdm}.
The deviation from the scaling behavior
for $(t-t_{\rm c})/R(t_{\rm c}) > 0.40$
seen in Fig.~\ref{fig:vdm10dp1point4_t_r2|eigen_a_vdm}
can be understood either as
the finite volume effects or as
the finite lattice spacing effects
depending on the parameter set.

\begin{figure}
\centering{}
%\begin{centering}
\includegraphics[scale=0.6]{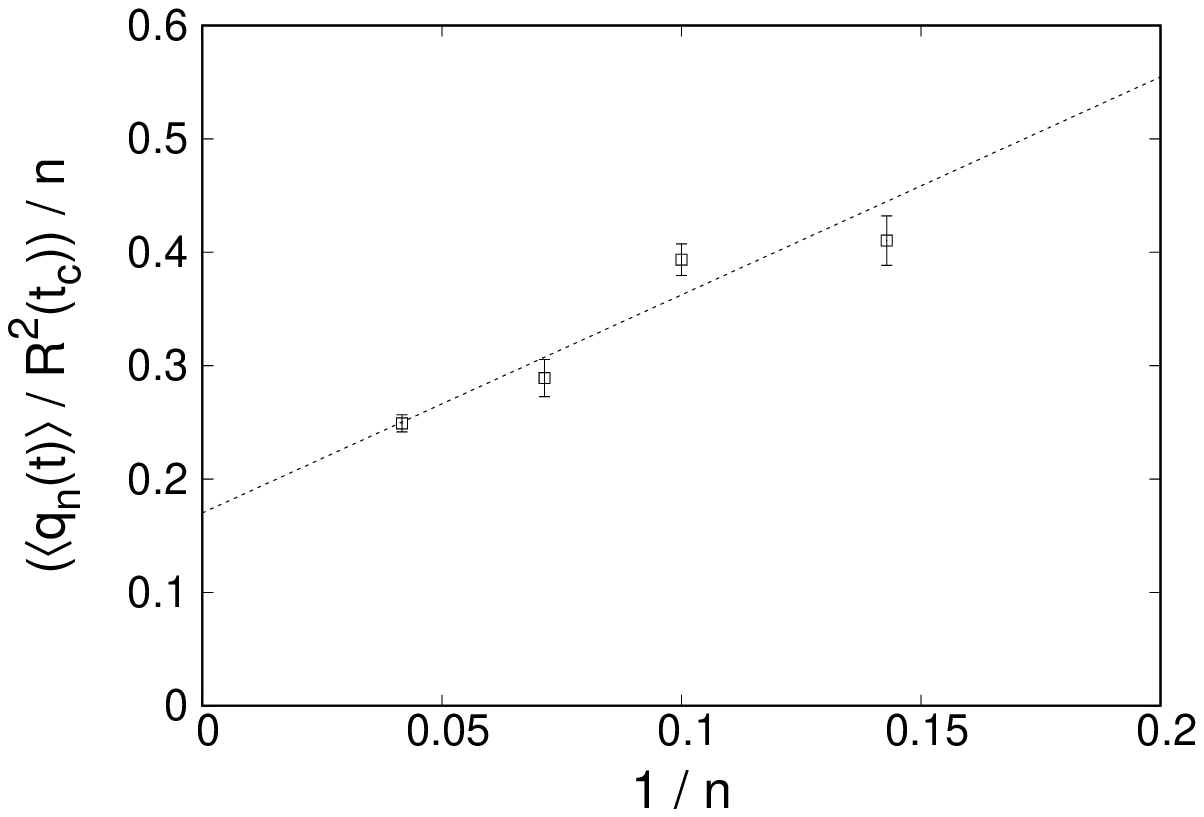}
\includegraphics[scale=0.6]{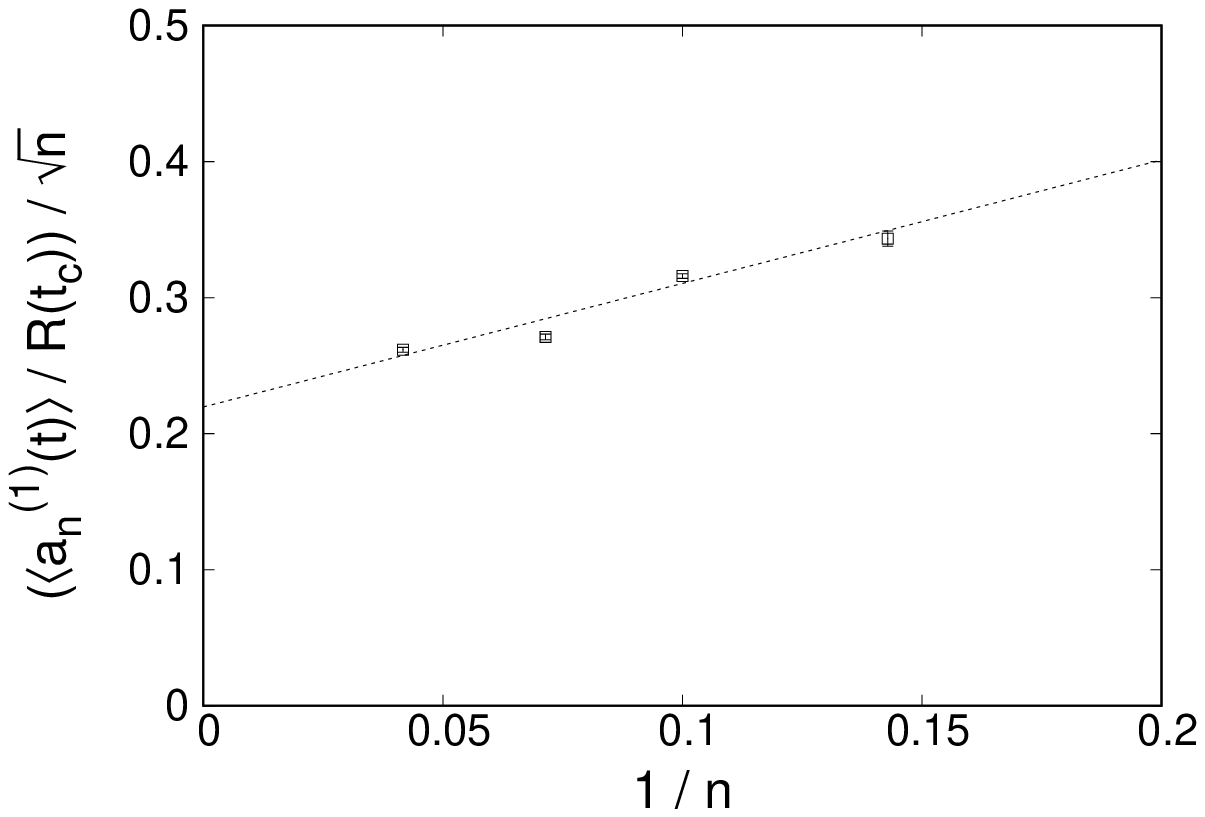}
%\par\end{centering}
%\centering{}
\caption{(Left) The largest eigenvalue $q_{n}(t)$
of the matrix $Q(t)$ 
obtained at $(t-t_{\rm c})/R(t_{\rm c}) \sim  0.40$
and normalized by $R^{2}(t_{\mathrm{c}})$ and $n$
is plotted against $1/n$.
(Right) The largest eigenvalue $a^{(1)}_{n}(t)$ of 
the matrix $\bar{A}^{(1)}(t)$
obtained at $(t-t_{\rm c})/R(t_{\rm c}) \sim  0.40$
and normalized by $R(t_{\mathrm{c}})$ and $\sqrt{n}$
is plotted against $1/n$.
}
\label{fig:qn-a1n_vdm} 
\end{figure}

In what follows, we focus on the point
$(t-t_{\rm c})/R(t_{\rm c}) \sim  0.40$,
at which the results for $R^2(t)/R^2(t_{\rm c})$
with the four parameter sets agree with each other.
%as one can see from 
%Fig.~\ref{fig:vdm10dp1point4_t_r2|eigen_a_vdm} (Top-Left).
%
%in Fig.~\ref{fig:vdm10dp1point4_t_r2|eigen_a_vdm} (Top-Left).
In Fig.~\ref{fig:vdm10dp1point4_t_r2|eigen_a_vdm} (Middle-Left),
we plot the normalized
eigenvalues $\langle q_{k}(t) \rangle/R^2(t_{\rm c})$ $(k=1,\cdots,n)$ of $Q(t)$
against their label $(k-1)/(n-1)$ for the four parameter sets.
This reveals a clear scaling behavior
except for the two largest eigenvalues,
which grow as the lattice spacing $\epsilon$ decreases.
This scaling behavior is consistent with the scaling of
the ratio $R^2(t)/R^2(t_{\rm c})$
%approcahes some constant
in the continuum limit \cite{Ito:2013ywa,Ito:2015mxa}
seen in the Top-Left panel
considering the relation (\ref{R2-q-eigen-rel}).
Note, however, that the time dependence of 
$R^2(t)/R^2(t_{\rm c})$ is caused by the two largest eigenvalues of $Q(t)$
as we have seen in the previous subsections.
Therefore, the scaling of $R^2(t)/R^2(t_{\rm c})$
implies that the two largest eigenvalues of $Q(t)$
should grow linearly 
in $n$ in the continuum limit.
This is confirmed numerically in Fig.~\ref{fig:qn-a1n_vdm} (Left)
assuming the presence of $1/n$ corrections.

%% In Fig.~\ref{fig:R2_bosonic|eigen_a_bosonic} (Bottom-Left),
%% we plot the eigenvalues $a^{(1)}_{k}(t)/R(t_{\mathrm{c}})$ against
%% $(t-t_{\mathrm{c}})/R(t_{\mathrm{c}})$.
%% We find that only
%% two eigenvalues $a_{1}^{(1)}(t)$ and $a_{n}^{(1)}(t)$ 
%% grow in magnitude with time $t$,
%% and all the others remain close to zero.
%% Similar behaviors are seen also for the eigenvalues
%% $a^{(2)}_{k}(t)$ and $a^{(3)}_{k}(t)$ obtained from the other 
%% two extended directions.

%% In Fig.~\ref{fig:R2_bosonic|eigen_a_bosonic} (Bottom-Right),
%% we plot the eigenvalues $a^{(4)}_{k}(t)/R(t_{\mathrm{c}})$ against
%% $(t-t_{\mathrm{c}})/R(t_{\mathrm{c}})$.
%% We find that all the eigenvalues
%% remain close to zero.
%% Similar behaviors are seen also for the eigenvalues
%% $a^{(2)}_{k}(t)$ and $a^{(3)}_{k}(t)$ obtained from the other 
%% five shrunken directions.
%% Comparing the results for the extended directions with those for
%% the shrunken directions,
%% we notice that the eigenvalue distribution of $\bar{A}^{(i)}$
%% is almost identical 
%% except for the two eigenvalues with the largest magnitude.
%% This implies that the spontaneous symmetry breaking of the SO(9) 
%% rotational symmetry is cause only by these two eigenvalues.

Let us next consider the space-time structure in the extended directions
and the shrunken directions separately.
In Fig.~\ref{fig:vdm10dp1point4_t_r2|eigen_a_vdm} (Middle-Right), 
we plot the eigenvalues 
of $\bar{A}^{(1)}(t)/R(t_{\rm c})$
%$\langle a^{(1)}_{k}(t) \rangle /R(t_{\mathrm{c}})$ 
obtained at $(t-t_{\rm c})/R(t_{\rm c})\approx0.40$
against the label $(k-1)/(n-1)$. 
Here again we observe a clear scaling behavior
except for the ones at both ends of the spectrum.
Similar behaviors are obtained
for the other extended directions.
%$\bar{A}_{2}(t)$ and $\bar{A}_{3}(t)$ as well as $\bar{A}_{1}(t)$. 
%% Let us recall that the the eigenvalues of $\bar{A}^{(i)}(t)$
%% % and those of $\bar{A}^{(i)}(t)$
%% are related to the extent of space $\lambda_i(t)$ 
%% as (\ref{lambda-a-eigen-rel}).
%% Note here that 
%% the ratio $\lambda_i(t)/R^2(t_{\rm c})$ is known to have a continuum 
%% limit \cite{Ito:2013ywa,Ito:2015mxa}.
%% This is consistent not only with
%% the scaling behavior observed above
%% but also with the growth of the two largest eigenvalues
%% % of $Q(t)$ normalized with $R^2(t_{\rm c})$
%% as the continuum limit is taken.
%% We can also deduce that the growth should be linear in $n$ in the continuum limit.
According to the same argument as in the previous paragraph,
we can deduce that the normalized eigenvalues at both ends of the spectrum
grow in magnitude as O($\sqrt{n}$) in the continuum limit,
which is confirmed in Fig.~\ref{fig:qn-a1n_vdm} (Right)
assuming the presence of $1/n$ corrections.

In Fig.\ \ref{fig:vdm10dp1point4_t_r2|eigen_a_vdm} (Bottom), 
we plot the eigenvalues 
of $\bar{A}^{(4)}(t)/R(t_{\rm c})$
%$a^{(4)}_{k}(t)/R(t_{\mathrm{c}})$
%corresponding to the shrunken direction
obtained at $(t-t_{\rm c})/R(t_{\rm c})\approx0.40$
%as the plot in the Middle-Right panel
against the label $(k-1)/(n-1)$.
%for the same time $(t-t_{\rm c})/R(t_{\rm c})\approx0.40$ as the plot on the left.
We observe a clear scaling behavior here as well.
In fact, the eigenvalues are almost the same as those
for the extended directions except for the ones at both ends.
Similar behaviors are obtained 
for the other shrunken directions.

%% This implies that the spontaneous breaking of the SO(9) rotational symmetry
%% is caused only by these 
%% Thus we find that the only difference between
%% the extended directions and the shrunken directions
%% appears at both ends of the eigenvalue spectrum of $\bar{A}^{(i)}(t)$.

Thus we find in the VDM model
that the singular space-time structure becomes even more pronounced
%instead of getting milder 
in the continuum limit
instead of getting milder.
It is surprising that 
%The fact that the spontaneous breaking of the SO(9) rotational symmetry
the two eigenvalues of $\bar{A}^{(i)}(t)/R(t_{\mathrm{c}})$ ($i=1,2,3$ )
%properly normalized by $$ 
actually diverges
in the continuum limit although 
the extent of space defined by $R^2(t)/R^2(t_{\rm c})$
%the eigenvalue distribution 
remains finite.
It is these two eigenvalues that
cause the spontaneous breaking of the SO(9) rotational symmetry
and the expansion of space.
All the other eigenvalues 
of $\bar{A}^{(i)}(t)/R(t_{\mathrm{c}})$
remain finite and contribute
only to the time-independent SO(9) symmetric part of 
the ``moment of inertia tensor'' $T_{ij}(t)$.
%(\ref{eq:def-tij}) 

%\section{The mechanism for the spontaneous symmetry breaking}
%\section{The Pauli-matrix structure}

%% two eigenvalues of $Q(t)$ become large 
%% we plot the normalized
%% eigenvalues $q_{k}(t)/R^2(t_{\rm c})$ $(k=1,\cdots,n)$ of $Q(t)$

%% \begin{figure}
%% \centering{}
%% \includegraphics[scale=0.6]{vdm_bulk_v3}
%% \includegraphics[scale=0.6]{vdm_Q}
%% \caption{(Left) The extent of bulk in each direction 
%% $\lambda_{i}^{\mathrm{bulk}}(t)/R(t_{\mathrm{c}})^{2}$
%% is plotted against the lattice spacing $\epsilon$ 
%% at time $(t-t_{\rm c})/R(t_{\rm c})\sim0.40$
%% for the VDM model. The parameters used to make this plot are presented
%% in table \ref{InformationInVDM}. 
%% (Right) The eigenvalues $q_{k}(t)/R(t_{\mathrm{c}})^{2}$
%% of the matrix $Q(t)$ defined in (\ref{eq.2}) are plotted against
%% $k/(n-1)$ for the same parameters as in the left figure, where $k$
%% labels the eigenvalues so that $q_{0}(t)<\cdots<q_{n-1}(t)$.}
%% \label{fig:lam_bulk_and_Q_eigen_vdm} 
%% \end{figure}

\subsection{The Pauli-matrix structure}
\label{sec:pauli-matrix}

In this subsection, we provide deeper understanding
of the singular space-time structure 
observed in the previous subsections.
%the mechanism of the emergence of (3+1)D 
%expanding space-time.
%Since $\bar{A}^{(i)}(t)$ are not simultaneously diagonalizable, 
Let us work 
in the SU($n$) basis which diagonalizes $Q(t)$ at each time $t$
with the ordering (\ref{q-ordering-def}),
and consider the $2 \times 2$ 
submatrix $X_i(t)$ in the bottom-right corner of
\begin{equation}
	\bar{A}^{(i)}(t)=
	\left(
		\begin{array}{cc}
			\ast & \ast \\
			\ast & X_i(t)
		\end{array}
	\right)
\label{def-X-block}
\end{equation}
for the extended directions $i=1,2,3$.
Here we use the VDM model 
with the parameter sets given in table~\ref{InformationInVDM}
and take the continuum limit 
focusing on the time $(t-t_{\rm c})/R(t_{\rm c})\approx0.40$
as we did in section \ref{sec:cont-lim}.

We show below that the three matrices $X_i$ in (\ref{def-X-block})
tend to satisfy the SU(2) Lie algebra
\begin{equation}
	[X_i, X_j] = i c \epsilon_{ijk}X_k
\label{SU2-Lie-alg}
%\ ,
\end{equation}
for some real constant $c$ in the continuum limit.
%This implies that the two large eigenvalues of $Q(t)$ and 
%$\bar{A}^{(i)}(t)$ are actually associated with the Pauli matrices.
In order to determine the optimal value of $c$, we consider
a quantity
\begin{equation}
S(c)\equiv {\rm tr}(i\epsilon_{ijk}[X_i, X_j]+2 c X_k)^2 \ ,
\end{equation} 
which represents the violation of the relation (\ref{SU2-Lie-alg}).
The value of $c$ that minimizes $S(c)$ can be readily obtained as
\begin{equation}
\tilde{c}
= - \frac{i\epsilon_{ijk}{\rm tr}(X_k[X_i, X_j])}{2{\rm tr}(X_l^2)} \ .
\label{tilde_c-def}
\end{equation} 
%Then we would like to see whether the relation (\ref{SU2-Lie-alg})
%is satisfied for $c=\tilde{c}$ to some accuracy.
Using $c=\tilde{c}$ as the optimal value for each configuration,
we investigate to what extent the relation (\ref{SU2-Lie-alg}) 
is satisfied.

\begin{figure}
\centering{}
\includegraphics[scale=0.48]{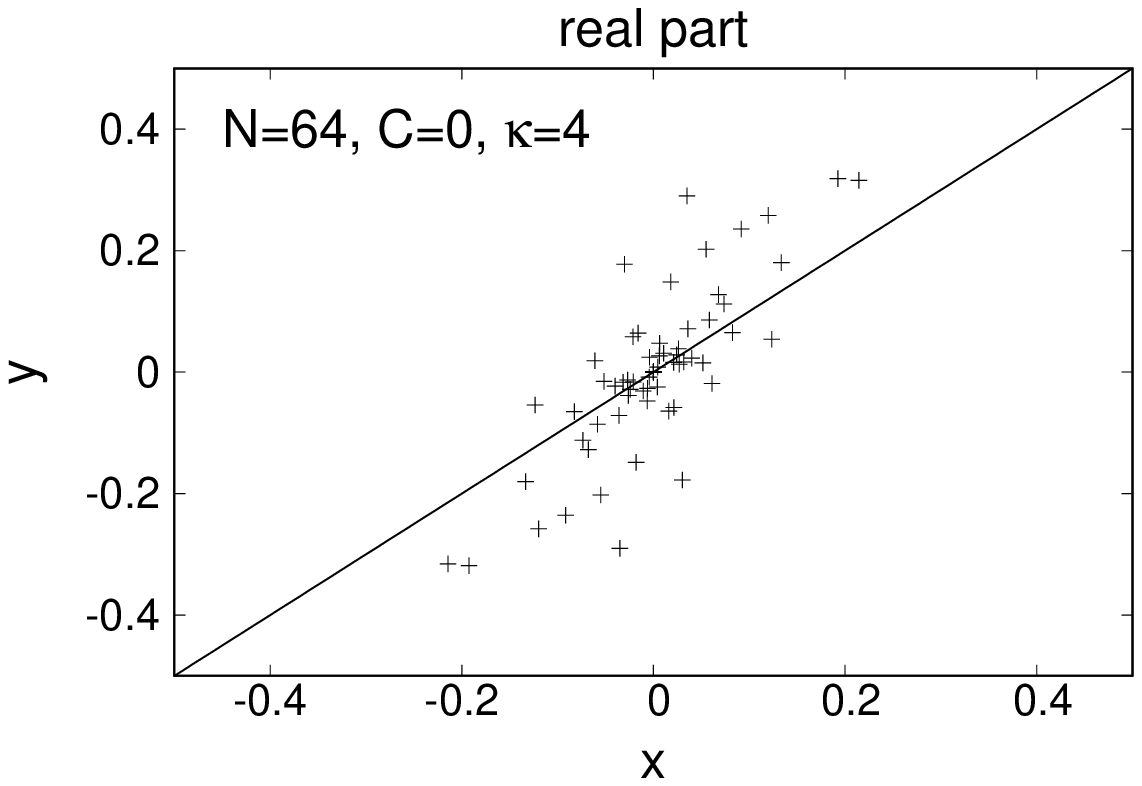}
\includegraphics[scale=0.48]{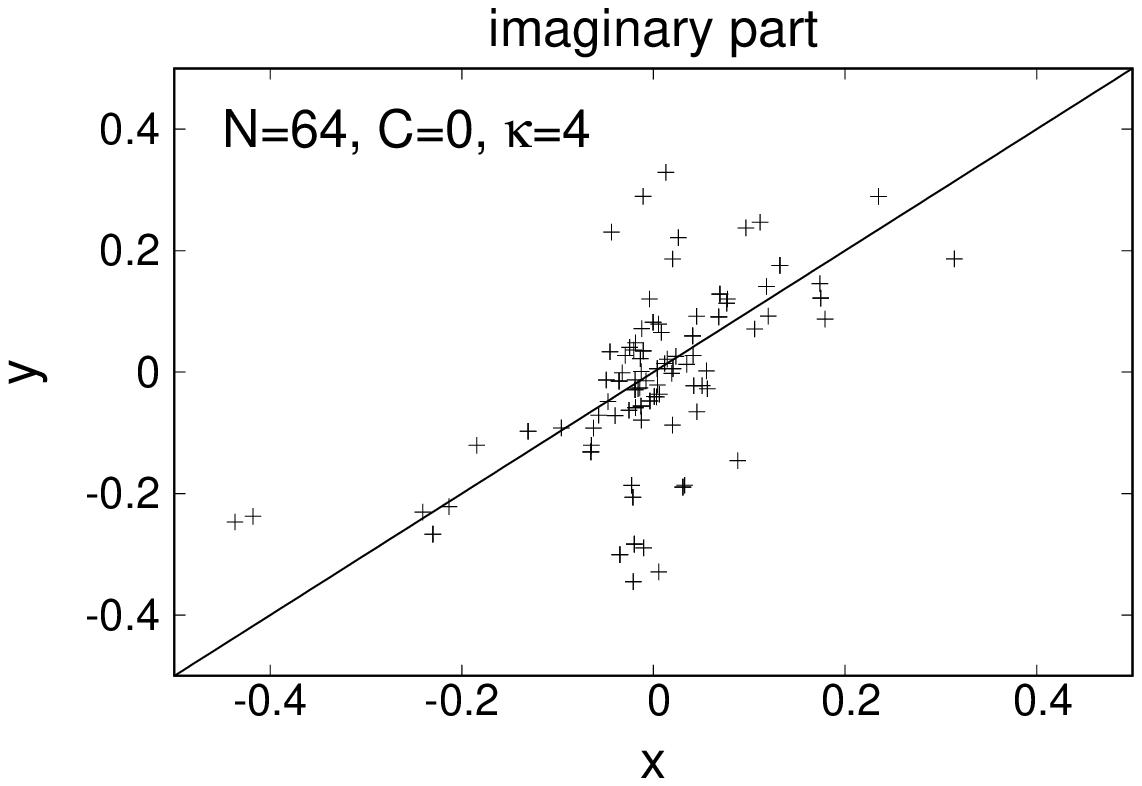}
\\
\includegraphics[scale=0.48]{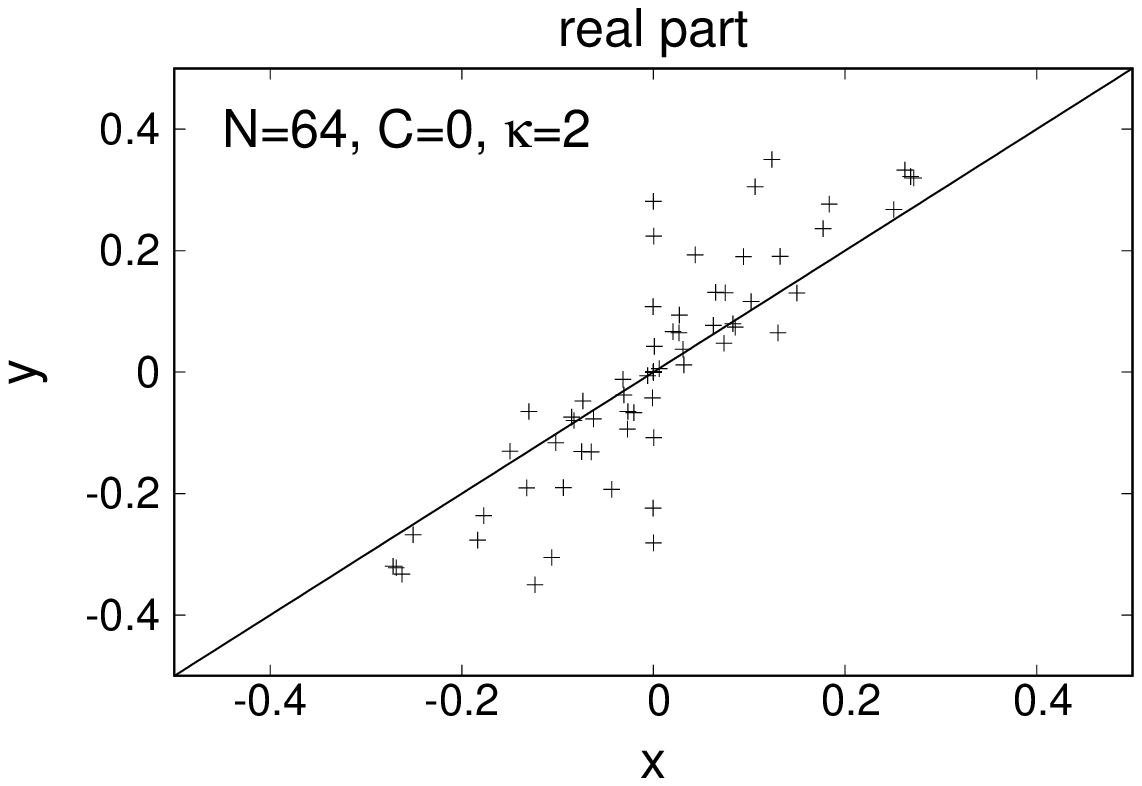}
\includegraphics[scale=0.48]{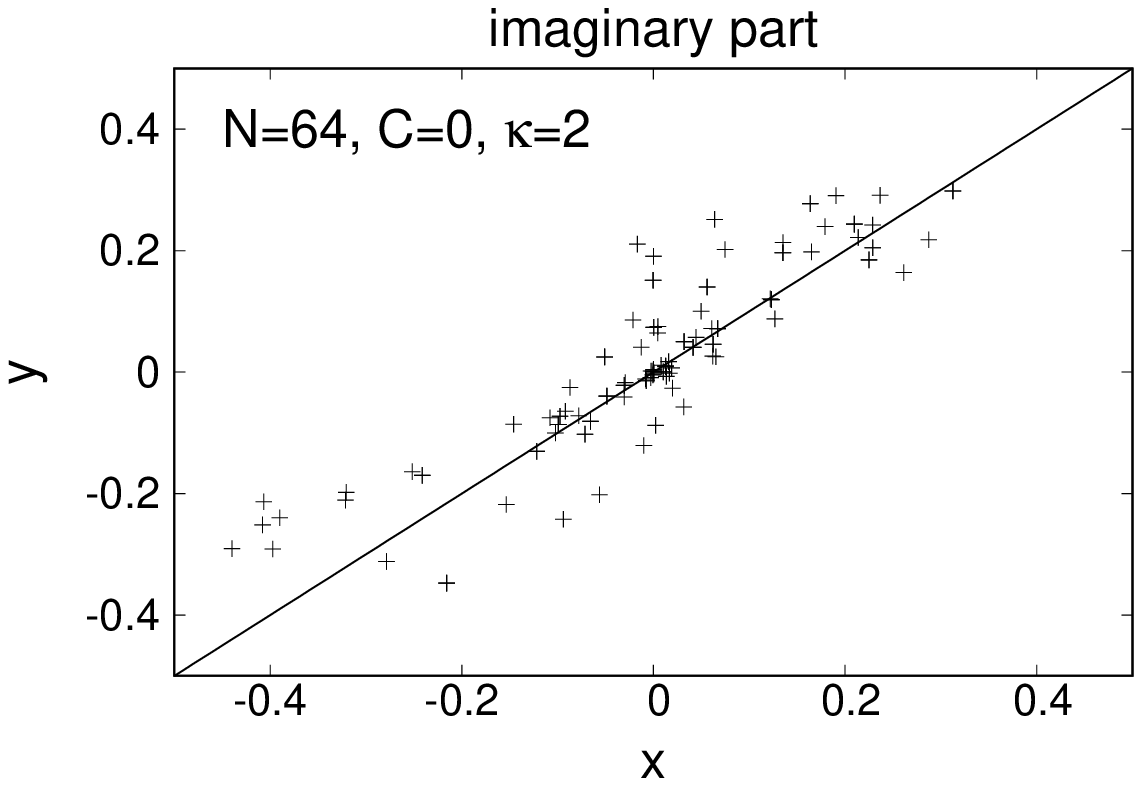}
%\vspace{5mm}
\\
\includegraphics[scale=0.48]{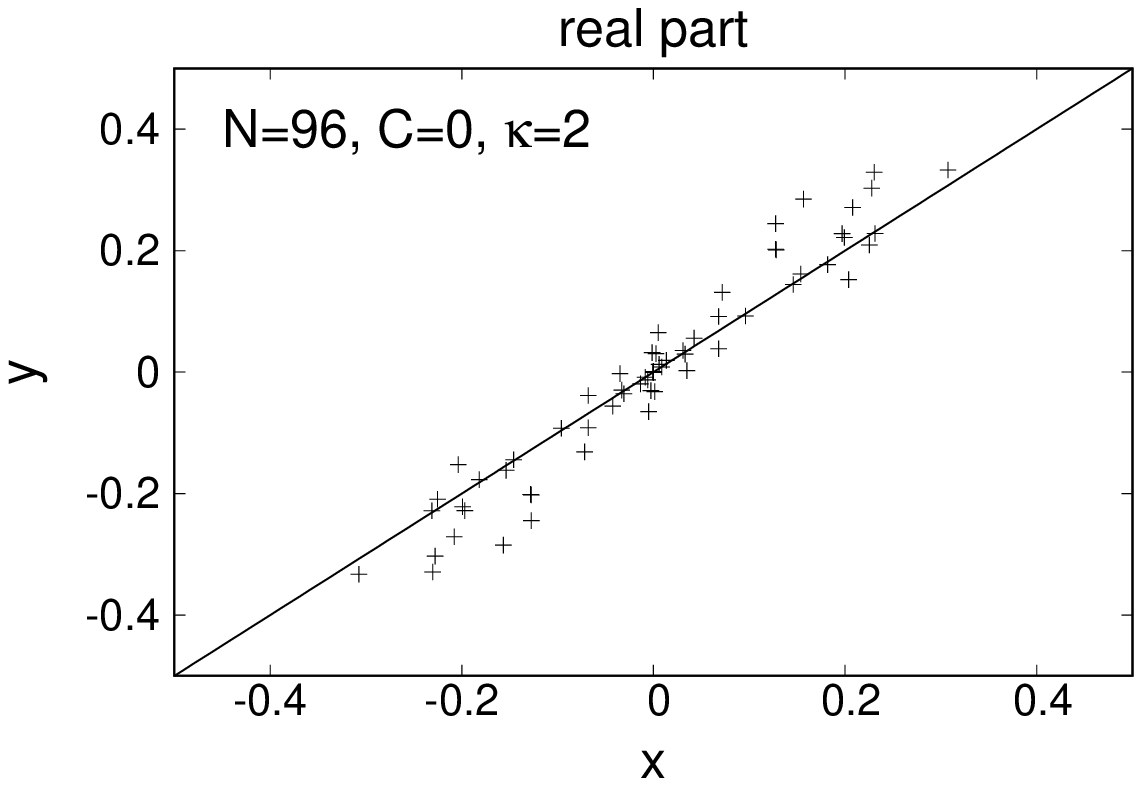}
\includegraphics[scale=0.48]{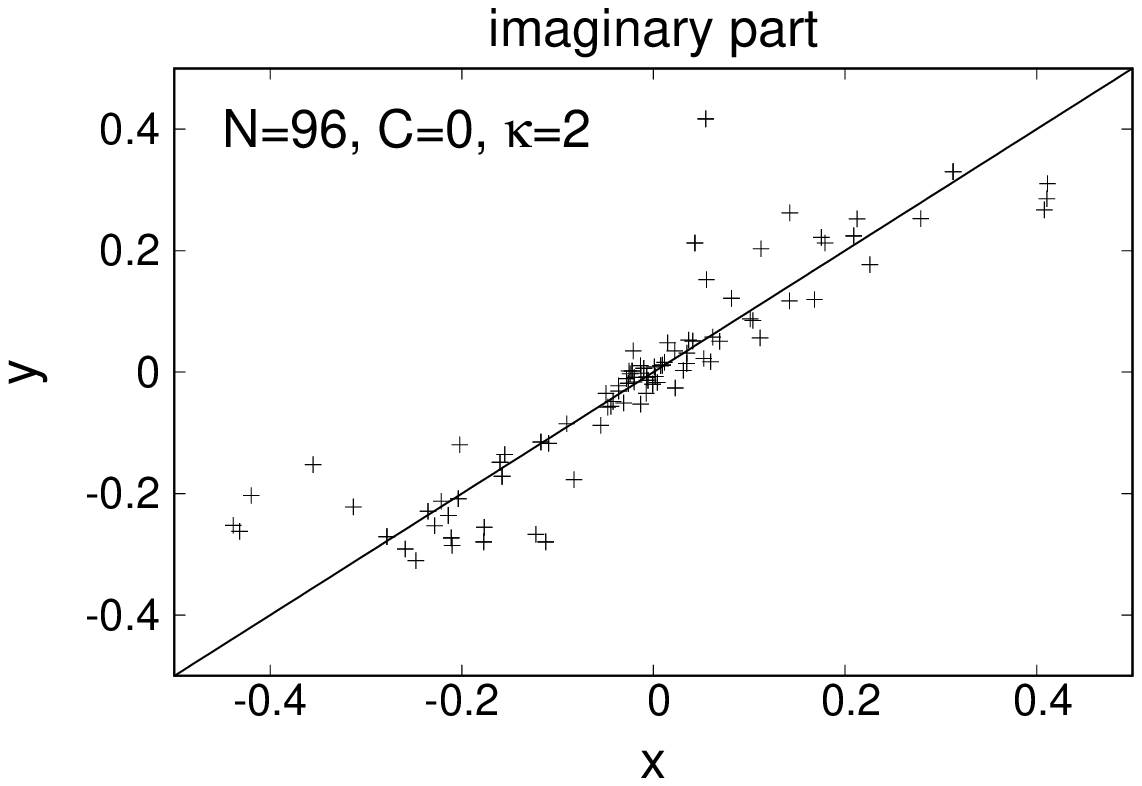}
%\vspace{5mm}
\\
\includegraphics[scale=0.48]{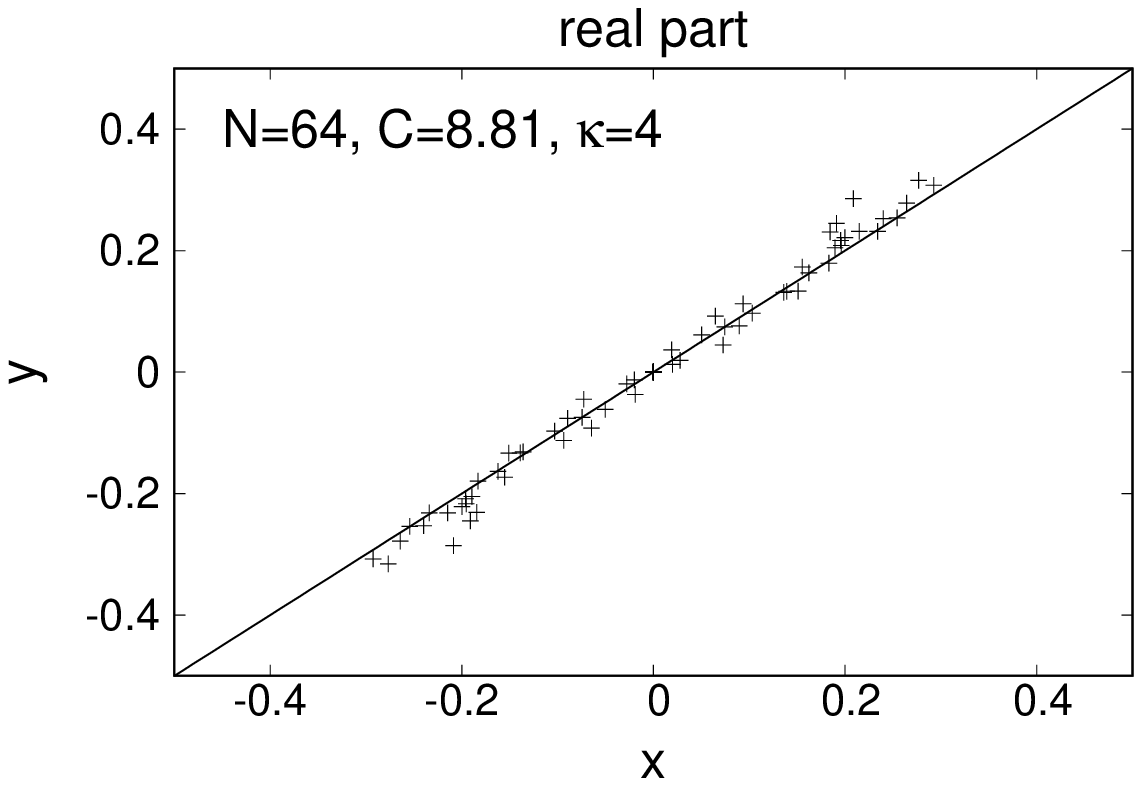}
\includegraphics[scale=0.48]{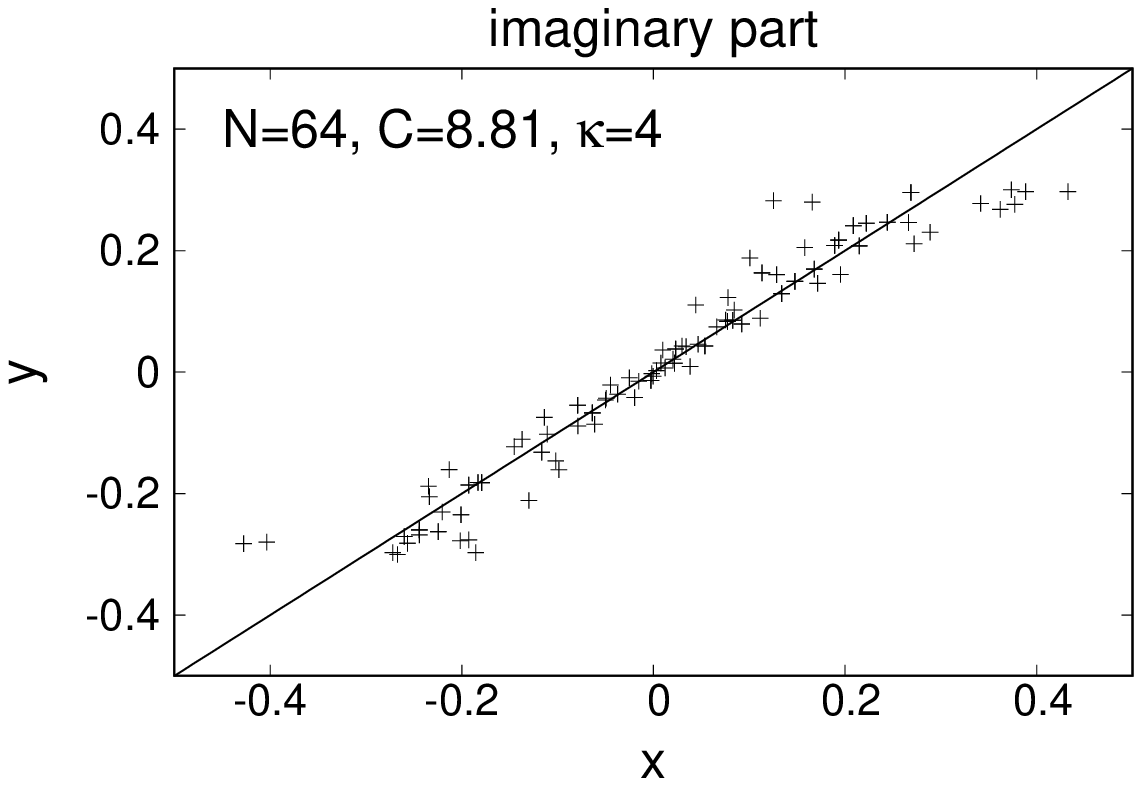}
\caption{(Left) A scatter plot for the real part
$x = {\rm Re} (i \tilde{c}\epsilon_{ijk}(X_k)_{ab})
/{\rm tr}(X_l^2)$
and 
$y={\rm Re}([X_i, X_j]_{ab})/{\rm tr}(X_l^2)$
of each side of (\ref{SU2-Lie-alg}) with (\ref{tilde_c-def})
is shown for $(i,j)=(1,2),(2,3),(3,1)$ and
$(a,b)=(1,1),(1,2),(2,2)$
using 10 configurations obtained by simulating
the VDM model 
with the parameter sets given in table~\ref{InformationInVDM}.
The solid line represents $y=x$.
(Right) A scatter plot for the imaginary part
$x = {\rm Im} (i \tilde{c}\epsilon_{ijk}(X_k)_{ab})
/{\rm tr}(X_l^2)$ 
and $y={\rm Im}([X_i, X_j]_{ab})/{\rm tr}(X_l^2)$
of each side of (\ref{SU2-Lie-alg}) with (\ref{tilde_c-def})
is shown in the same way.
}
	\label{algebra}
\end{figure}

In Fig.~\ref{algebra},
we show a scatter plot
for the real part (Left)
and the imaginary part (Right)
of each side of (\ref{SU2-Lie-alg}).
%with (\ref{tilde_c-def}).
%with the coefficient chosen as $c=\tilde{c}$ for each configuration.
The quantities on both sides
are normalized by ${\rm tr}(X_l^2)$
%for fair comparison between different parameter sets.
so that they become invariant under the scale transformation 
$X_i \mapsto {\rm const.} X_i$.
We observe that 
the data points tend to 
converge to the line $y=x$ as 
one goes from the top to the bottom
corresponding to decreasing
the lattice spacing $\epsilon$ (See table~\ref{InformationInVDM}.).
%% Similar behaviors are observed also by using the imaginary part
%% instead of the real part.
This shows that the $2\times 2$ matrices
$X_i$ ($i=1,2,3$)
tend to 
satisfy (\ref{SU2-Lie-alg})
%have a Pauli-matrix struture
in the continuum limit.

Thus we conclude that the singular space-time structure
observed for the matrix configurations generated by simulations
is essentially associated with the Pauli matrices.
The Pauli matrices may be regarded as the simplest matrix configuration
that has SO(3) symmetry in the sense that their SO(3) rotation can be
absorbed by an appropriate SU($N$) transformation.
Given the situation 
characterized by the two large eigenvalues of $Q(t)$, 
the appearance of the Pauli-matrix structure may not be that surprising.

%% Let us discuss the approach to the Pauli-matrix structure
%% more quantitatively.
%% For that, we define
%% \begin{equation}
%% \delta \equiv \frac{S(\tilde{c})}{({\rm tr}(X_i^2))^2} \ ,
%% \end{equation}
%% which vanishes
%% if and only if $X_i$ satisfy the SU(2) Lie algebra.
%% The denominator makes $\delta$
%% invariant under the scale transformation 
%% $X_i \mapsto {\rm const.} X_i$.
%% In fig.~\ref{dif_Pauli}, we plot the expectation value
%% $\langle \delta \rangle$ for each parameter set
%% in table~\ref{InformationInVDM}
%% against $1/n$, where $n$ is the block size.
%% This suggests that $\langle \delta \rangle$ vanishes 
%% in the $n\rightarrow \infty$ limit as $O(1/n)$.

%% \begin{figure}
%% \begin{center}
%% 	\includegraphics[width=8cm]{pauli.eps}
%% \caption{The expectation value of $\delta$ is 
%% plotted against $1/n$ for the VDM model
%% for the parameter sets in table~\ref{InformationInVDM}. 
%% The solid line is a fit to $y=ax$ with $a=4.28(37)$.}
%% \label{dif_Pauli}
%% \end{center}
%% \end{figure}

%\section{The approximation used in Monte Carlo simulation}
\section{The new interpretation of the simulation}
\label{sec:approximation}

%In this appendix we review the derivation of 
%eq.~(\ref{our-model}) following the original paper \cite{KNT}.

In this section, we attribute the observed Pauli-matrix structure
%observed in the previous section
to the approximation involved in deriving
the partition function (\ref{eq:partition-function2}),
which was used in Monte Carlo simulation.
%which was necessary 
%used in Monte Carlo simulation.
We point out a subtlety in the approximation, 
and argue that the approximation amounts
to replacing $\ee^{iS_{\mathrm{b}}}$ by $\ee^{\beta S_{\mathrm{b}}}$
in the original partition function (\ref{eq:partition-function}).
This new interpretation of the simulation 
provides us with a natural understanding of
the (3+1)d expanding behavior
with the Pauli-matrix structure discussed in section
\ref{sec:space-time-structure}.
We also speculate on a possible scenario for the original model
with the correct $\ee^{iS_{\mathrm{b}}}$ factor.

%% The partition function (\ref{part-lorentzian})
%% is not finite as it stands,
%% but it can be made finite
%% by introducing infrared cutoffs 
%% in both temporal and spatial directions as \cite{KNT}
%% \begin{eqnarray}
%% %\frac{1}{N}\mbox{Tr}\left(A_{0}\right)^{2}
%% %& \leq & \kappa \, L^{2} \ ,
%% \frac{1}{N}\mbox{Tr}\left(A_{0}\right)^{2}  
%% & \leq &  \kappa \frac{1}{N}\mbox{Tr}\left(A_{i}\right)^{2}  \ ,
%% \label{kappa}\\
%% \frac{1}{N}\mbox{Tr}\left(A_{i}\right)^{2} & \leq & \Lambda^{2} \ .
%% \label{L}
%% \end{eqnarray}
%% It turned out that these two cutoffs can be
%% removed in the large-$N$ limit in such a way
%% that physical quantities scale.
%% %(such as the extent of space) 
%% The resulting theory thus obtained
%% has no parameter except one scale parameter.

\subsection{The ``derivation'' of 
the partition function (\ref{eq:partition-function2})}

Let us first review how one can obtain
the partition function (\ref{eq:partition-function2})
used in Monte Carlo simulation
from the original partition function (\ref{eq:partition-function}).
(This was done in Appendix A of ref.~\cite{Ito:2013ywa} for $p=1$,
but here we generalize it to arbitrary $p$.)

Note
that the integrand of the partition function (\ref{eq:partition-function})
involves a phase factor $\ee^{iS_{\rm b}}$.
%just as in the path-integral formulation of quantum field theories
%in Minkowski space.
As is commonly done in integrating oscillating functions,
we introduce the convergence factor
$\ee^{- \epsilon |S_{\rm b}|}$ and take the $\epsilon \rightarrow 0$ limit
after the integration.

The partition function can then be rewritten as
\begin{align}
Z &= \int dA \int_0^{L^{2p}} d(r^p) \,
\delta\left(
\frac{1}{N}\mbox{Tr}\left\{ (A_{i})^{2} \right\}^p - r^p \right)
\theta\left(\kappa^p r^p - \frac{1}{N}\Tr (A_0)^{2p}  \right) 
 \ee^{i S_{\rm b}-\epsilon |S_{\rm b}| }
\, {\rm Pf} {\cal M} \  ,
\label{Z-rewritten}
\end{align}
where $\kappa$ and $L$ are the cutoff parameters
introduced in (\ref{eq:cutoff-temp}) and 
(\ref{eq:cutoff-space}), respectively.
%% where the integration over $A_\mu$ is assumed to be
%% restricted by
%% the constraint (\ref{kappa}).
Rescaling the variables $A_\mu \mapsto r^{1/2} A_\mu$
in the integrand,
%, so that the integration over $A_\mu$ is now
%restricted to $\frac{1}{N}\tr (A_i)^2=1$ due to
%the inserted delta-function.
%we integrate over $r$ and 
we get
\begin{align}
Z = & \int dA \,
{\rm Pf} {\cal M} (A) \,
%\delta\left(
f(S_{\rm b}) \,
% \nonumber \\
%& \times
\delta\left(\frac{1}{N}\Tr \{(A_i)^2\}^p - 1 \right)
\theta\left(\kappa^p  - \frac{1}{N}\Tr (A_0)^{2p}  \right)  \ .
\label{our-model2}
\end{align}
Here we have defined the function $f(S_{\rm b})$ by
%when we go from the first line to the second
%Integration over $r$ can be performed and one obtains
\beq
f(S_{\rm b}) \equiv
\int_0^{L^{2p}}  d(r^p)  \, r^{9(N^2-1)-1}
 \ee^{r^{2} (i S_{\rm b}-\epsilon |S_{\rm b}|) }
 \ ,
\label{integrate-r}
\eeq
%% If it were not for the cutoff (\ref{L}),
%% we have $L=\infty$ and $f(S_{\rm b})$ diverges
%% severely at $S_{\rm b}=0$ as
%% \beq
%% f(S_{\rm b})
%% \propto \frac{1}{|S_{\rm b}|^{\frac{18}{4}(N^2-1)}} \ .
%% \eeq
%% For finite $L$, the function $f(S_{\rm b})$ is finite,
%% but it still has a sharp peak at $S_{\rm b}=0$.
which 
%Note that $f(S_{\rm b})$
is a complex-valued function with the property 
$f(-S_{\rm b}) = f(S_{\rm b})^{*}$.

For $|S_{\rm b}| \ll \frac{1}{L^4}$,
the function can be well approximated by 
\beq
f(S_{\rm b}) \approx 
\frac{p}{9(N^2-1)+p-1} (L^2)^{9(N^2-1)+p-1} \ .
\label{f-Sb-small}
\eeq
For $|S_{\rm b}| \gtrsim \frac{1}{L^4}$,
on the other hand,
the phase of the integrand in (\ref{integrate-r})
starts to oscillate violently in the region 
$r \gtrsim 1/\sqrt{|S_{\rm b}|}$,
and hence the integral decreases rapidly in magnitude
for increasing $|S_{\rm b}|$.
In particular, 
the asymptotic behavior 
of $f(S_{\rm b})$
for $S_{\rm b} \gg \frac{1}{L^4}$ can be 
estimated as
\beq
%\frac{f(S_{\rm b})}{f(0)} 
\frac{|f(S_{\rm b})|}{f(0)} 
%\sim
%\approx
=
%e^{i\frac{\pi}{4} \{ 9(N^2-1) + p - 1 \}  }
\Gamma \left( \frac{9(N^2-1)+p+1}{2}  \right)
\,
\left( 
\frac{1}{L^4 |S_{\rm b}|} \right)^{\frac{9(N^2-1)+p-1}{2}} 
+ O (\ee^{-\epsilon L^4 |S_{\rm b}| })
%% f(S_{\rm b}) \sim \{ 9(N^2-1)-1 \} ! 
%% \left( 
%% \frac{1}{|S_{\rm b}|} \right)^{\frac{9(N^2-1)}{2}} \ ,
%\frac{1}{9(N^2-1) (|S_{\rm b}|)^{9(N^2-1)/2}} \ .
\eeq
by deforming the integration contour in (\ref{integrate-r}).
%and using the asymptotic behavior of the incomplete gamma function.

Recalling eq.~(\ref{eq:sb2}),
the condition $|S_{\rm b}| \ll \frac{1}{L^4}$
for (\ref{f-Sb-small}) can be rewritten as
\beq
\left| \frac{1}{N}\Tr (F_{\mu\nu}F^{\mu\nu})\right|  \ll 
%\frac{4 g^2}{N L^4} \ .
\frac{4}{N L^4} \ .
\label{F2-condition}
\eeq
Therefore, assuming that the right-hand side 
%$\frac{4 g^2}{N L^4}$ 
$\frac{4}{N L^4}$ 
of (\ref{F2-condition})
becomes small at large $N$,
we may make a replacement
\beq
f(S_{\rm b}) 
\Longrightarrow
%\mapsto 
\delta\left(\frac{1}{N}\Tr (F_{\mu\nu}F^{\mu\nu})  \right)
\label{eq:replacement}
\eeq
up to a normalization constant.
%% In fact, allowing $S_{\rm b}$ to have a small value does not make
%% much difference since 
%% Here we note that $S_{\rm b}$ is a difference of two
%% positive definite terms (\ref{F2-two-terms}),
%% and therefore 
%% allowing $|S_{\rm b}|$ to have a value much smaller than
%% typical values taken by the two terms 
%% in the path integral
%% %in (\ref{F2-two-terms})
%% hardly makes any difference.
%% This can be checked explicitly by actual Monte Carlo simulation.
For 
%Similar ``derivations'' can be applied to 
the bosonic model and the VDM model,
% (\ref{vdm-model}).
one simply has to replace the Pfaffian
in (\ref{Z-rewritten}) and (\ref{our-model2}) as
(\ref{eq:def-vdm-bosonic}).

\subsection{Subtlety in the derivation and the new interpretation}

The only step in the derivation that may go wrong 
is the replacement (\ref{eq:replacement}).
The subtlety in this replacement can be seen as follows.
Note that the phase factor $\ee^{iS_{\rm b}}$
in the partition function (\ref{eq:partition-function})
favors configurations at which the bosonic action $S_{\rm b}$ 
is stationary. On the other hand, 
the above approximation essentially replaces
the phase factor $\ee^{iS_{\rm b}}$ by the delta function
$\delta (S_{\rm b})$, which amounts to 
picking up configurations at which $S_{\rm b}$ is stationary only under 
rescaling $A_\mu \mapsto {\rm const.} A_\mu$.
While it is true that $|f(S_{\rm b})|$ is sharply peaked
at $S_{\rm b}=0$, the function $f(S_{\rm b})$ is actually 
a complex-valued function, whose phase rotates violently
around $S_{\rm b} = 0$. This effect of the phase should be
responsible for favoring the configurations at which 
$S_{\rm b}$ is stationary.
The approximation ignores this effect completely,
and hence it cannot be justified.
%it is not surprising that the stationary configurations 
%are no longer favored.

If the model (\ref{eq:partition-function2}) is not equivalent
to the original model (\ref{eq:partition-function}),
what kind of model does it actually correspond to?
Here we point out that 
the constraint on $S_{\rm b}$ that appears in
(\ref{eq:partition-function2})
may be regarded as the constraint one uses in defining
a microcanonical ensemble.
From this viewpoint, we consider that
the model (\ref{eq:partition-function2}) is actually 
equivalent to the corresponding canonical ensemble
with the Boltzmann weight $\ee^{\beta S_{\rm b}}$.
The real parameter $\beta$ depends on the parameter $C$
in the constraint\footnote{This connection also
provides clear justification
of the renormalization-group-like 
method \cite{Ito:2013ywa,Azuma:2017dcb},
which amounts to tuning the parameter $C$ in order to
obtain the late-time behaviors with smaller matrix size.}.
As we will see below, 
%from the behaviors observed in the simulation,
we consider that 
the model (\ref{eq:partition-function2})
%our Monte Carlo simulation as well as the previous ones 
corresponds essentially 
to replacing $\ee^{iS_{\rm b}}$ by $\ee^{\beta S_{\rm b}}$
with $\beta > 0$.

For $\beta > 0$, the first term in (\ref{eq:sb2}) that appears
in $\ee^{\beta S_{\rm b}}$ favors configurations
in which $A_0$ and $A_i$ commute.
This means that the spatial matrices $A_i$ tend to become diagonal
in the SU($N$) basis which diagonalizes $A_0$.
On the other hand, the second term in (\ref{eq:sb2}) favors configurations
in which the noncommutativity among the spatial matrices $A_i$ is large.
The band-diagonal structure, which plays a crucial role in 
extracting the real-time evolution as in 
section \ref{sec:time-evolution},
can be understood as
a consequence of the balance of these two effects.

We can also understand
the reason for the (3+1)d expanding behavior
with the Pauli-matrix structure.
Here we assume that the first term in (\ref{eq:sb2}) is 
not important except in realizing the band-diagonal structure
and focus on the effect of the second term in (\ref{eq:sb2}), which
favors large $\mathrm{Tr}\left(F_{ij}\right)^{2}$,
where $F_{ij}=i\left[A_{i},A_{j}\right]$. 
We also have to take into account the constraint 
$\frac{1}{N}\mathrm{Tr}\left\{\left( A_{i}\right)^{2}\right\}^p=1$,
where we set $p=1$ in what follows.
%noncommutativity among the spatial matrices $A_i$.

%the mechanism for the SSB of SO(9) rotational symmetry,
%which has been speculated only by a hand-waving argument 
%in ref.~\cite{Kim:2011cr}.
Simplifying the band-diagonal structure
of the spatial matrices $A_i$ ($i=1, \cdots , 9$),
we consider the block-diagonal structure given as
\begin{equation}
 A_i  = 
\left(
\begin{array}{cccc}
\bar{A}^{(1)}_i &  &  &   \\
                & \bar{A}^{(2)}_i &  &   \\
                &  &  \ddots  &   \\
                &  &  &  \bar{A}^{(B)}_i
\end{array}
\right) \ ,
\label{A_i-simplified}
\end{equation}
where $n$ is the common block size and $B$ is the number of blocks
satisfying $N=nB$.
%% We denote the $n \times n$ block matrices 
%% $\bar{A}^{(b)}_i$, where $b=1, \cdots , B$ 
%% is the label for each block. 
Within this ansatz,
we would like to maximize $\Tr (F_{ij})^2$
under the constraint $\frac{1}{N} \Tr (A_i)^2 = 1$.
Note that we have
\begin{align}
\frac{1}{N} \Tr (A_i)^2 & = \frac{1}{B} \sum_{b=1}^{B}  
\frac{1}{n} \Tr (\bar{A}^{(b)}_i)^2  \ , \\
\frac{1}{N} \Tr (F_{ij})^2 & = \frac{1}{B} \sum_{b=1}^{B}  
\frac{1}{n} \Tr (\bar{F}^{(b)}_{ij})^2   \ ,
\end{align}
where we have defined 
$\bar{F}^{(b)}_{ij}= i [\bar{A}^{(b)}_i , \bar{A}^{(b)}_j]$
for each block $b$.

Let us solve the maximization problem in two steps.
First we fix 
\begin{align}
\frac{1}{n} \Tr (\bar{A}^{(b)}_i)^2 & = (r_b)^2 \ , 
\\
\quad \frac{1}{B} \sum_{b=1}^B (r_b)^2 &= 1 \ ,
\label{rb_2_sum}
\end{align}
and maximize $\Tr (F_{ij})^2$ under this constraint.
Following the discussion given in ref.~\cite{Kim:2011cr},
the solution to this first maximization problem
can be written in terms of the Pauli matrices $\sigma_i$ as
\begin{align}
\bar{A}^{(b)}_i = \frac{1}{\sqrt{6}}
 r_b ( \sigma_i \oplus {\bf 0}_{n-2}) \ ,
\label{block-pauli}
\end{align}
for $i=1,2,3$ and $\bar{A}^{(b)}_i = 0 $ otherwise,
%where $\sigma_i$ $(i=1,2,3)$ are Pauli matrices,
up to the symmetries of the problem such as 
the SO(9) rotational symmetry
and the SU($n$) symmetry within each block.
The value of $\Tr (F_{ij})^2$ for (\ref{block-pauli})
is given as
\begin{align}
\Tr (F_{ij})^2 & = \frac{2}{3} \sum_{b=1}^{B}  
(r_b)^4   \ .
\label{f_pauli}
\end{align}
As the second step of the maximization, we
maximize (\ref{f_pauli}) under the constraint (\ref{rb_2_sum}).
The maximum is given when all but one of the $r_b$'s are zero.
%and the others are zero.

%% Taking the derivative with respect to $\bar{A}^{(b)}_i$, 
%% we get
%% \begin{align}
%% 2 [ \bar{A}^{(b)}_j , [ \bar{A}^{(b)}_j , \bar{A}^{(b)}_i ] 
%% - \lambda \bar{A}^{(b)}_i = 0 \ .
%% \end{align}

In reality, one should also take into account the entropic factor
due to quantum fluctuations, which is expected to favor
certain distribution of $r_b$.
% as well as deviation from
%the strict Pauli-matrix structure (\ref{block-pauli}).
Due to the time-reversal symmetry $A_0 \mapsto - A_0$ 
of the model, the most natural distribution would be
that $r_b$ is large around $t=0$ and decreases with $|t|$.
Thus we can understand the appearance of the (3+1)d
expanding behavior with the Pauli-matrix structure.

\subsection{A possible scenario for the original model}

In the previous subsections, 
we have argued that 
the model (\ref{eq:partition-function2}) used for
Monte Carlo simulation
actually corresponds to a model with $\ee^{\beta S_{\rm b}}$
instead of $\ee^{iS_{\rm b}}$ 
in (\ref{eq:partition-function}).
This new interpretation explains naturally 
the (3+1)d expanding behavior
with the Pauli-matrix structure.
%of the matrix configurations generated by simulation 
%if that is the case.
The crucial question then is what happens 
for the model with the correct $\ee^{iS_{\rm b}}$ factor.
It is not easy to answer this question due to the sign problem,
which occurs because 
$\ee^{iS_{\rm b}}$ is a pure phase factor and one cannot 
regard the integrand of the partition function 
(\ref{eq:partition-function}) as the probability distribution.
Here we speculate on a possible scenario based on
the results obtained so far.

For that purpose, let us consider a generalized model
with a factor
$\ee^{\beta (\cos\theta  + i \sin \theta) S_{\rm b}}$ 
($0 \le \theta \le \pi/2$),
which interpolates the two models.
At $\theta=0$, we obtain
the model 
with the positive definite factor $\ee^{\beta S_{\rm b}}$
we have been studying, whereas
at $\theta=\pi/2$,
we obtain the model with $\ee^{i\beta S_{\rm b}}$
we are aiming at.
The scale parameter $\beta$ can be absorbed, if one wishes,
by the redefinition $A_\mu \mapsto \beta^{-1/4} A_\mu$
and the replacement 
$L \mapsto \beta^{1/4} L$ in (\ref{eq:cutoff-space}).

As far as $\theta < \pi/2$, the real part of the coefficient of 
$S_{\rm b}$ is positive.
Therefore, certain effects
favoring the band-diagonal structure and the Pauli-matrix structure
in $A_i$ are at work. 
Note also that the classical equation of motion
% has a term 
%$\frac{\delta S_{\rm b}}{\delta A_\mu} $
%%%%%%%%coming from the bosonic part of the action 
%in common 
is common
to all values of $\theta$.
In fact, the classical equation of motion becomes valid
%for sufficiently large $\beta$, and even for any $\beta$,
%it becomes valid 
at late times if the expansion of space occurs
because each term in the bosonic
action becomes large \cite{Kim:2011ts,Kim:2012mw}.
Therefore, if some classical solution dominates
for $\theta=0$, the same solution may well dominate also for 
other $\theta$ less than some value $\theta_0$.
From this argument, we speculate that the 
models with $0\le \theta \le \theta_0$ are qualitatively the same.

As one approaches $\theta = \pi/2$,
the real part of the coefficient of $S_{\rm b}$ becomes small,
and different classical solutions may dominate.
Note that the matrix configurations with
the Pauli-matrix structure are obtained essentially
by maximizing $S_{\rm b}$, but 
the classical solutions that can be obtained
by \emph{extremizing} $S_{\rm b}$ instead of \emph{maximizing} it
should have more variety.
Indeed we have generated numerically 
many classical solutions
% with the IR cutoff 
that have (3+1)d expanding behavior  
% \emph{without} the Pauli-matrix structure; namely 
and find for all of them that
the matrix $Q(t)$ defined in (\ref{eq.2})
has a smooth eigenvalue distribution \cite{classical}.
This is understandable since the configurations
with the Pauli-matrix structure 
are actually disfavored entropically.
Recall, for instance, that only two eigenvalues of the matrix $Q(t)$ 
are large, meaning that the entropy for such configurations must be 
small.
It should be mentioned, however, 
that from the above classical analysis alone,
one cannot single out the (3+1)d expanding space-time
because there are also other solutions with different dimensionality.
Whether
the (3+1)d expanding behavior remains even for $\theta \sim \pi/2$
is therefore a highly nontrivial question.

%% It is therefore possible that as soon as 
%% the Pauli-matrix structure 
%% However, if one considers
%% the full quantum theory 
%% and approaches $\theta =\pi/2$ 
%% in the large-$N$ limit,
%% one may obtain
%% a regular space-time with the (3+1)d expanding behavior.

%% Some numerical evidence supporting this speculation shall be 
%% reported in a separate paper \cite{workinprog},
%% where the sign problem for $\theta \neq 0$
%% %caused by the phase factor $\ee^{i\beta S_{\rm b}}$
%% is overcome by the complex Langevin method.

%%%%%%%%%%%%%%%%%%%%%%%%%%%%%%%%%%%%%%%%%%%%%%%%%%%%%%%%%%%%%%%%%%%%%%
%%%%%%%%%%%%%%%%%%%%%%%%%%%%%%%%%%%%%%%%%%%%%%%%%%%%%%%%%%%%%%%%%%%%%%
%%%%%%%%%%%%%%%%%%%%%%%%%%%%%%%%%%%%%%%%%%%%%%%%%%%%%%%%%%%%%%%%%%%%%%\
\section{Summary and discussions}
\label{sec:summary}
%%%%%%%%%%%%%%%%%%%%%%%%%%%%%%%%%%%%%%%%%%%%%%%%%%%%%%%%%%%%%%%%%%%%%%
%%%%%%%%%%%%%%%%%%%%%%%%%%%%%%%%%%%%%%%%%%%%%%%%%%%%%%%%%%%%%%%%%%%%%%
%%%%%%%%%%%%%%%%%%%%%%%%%%%%%%%%%%%%%%%%%%%%%%%%%%%%%%%%%%%%%%%%%%%%%%\

In this paper we have investigated the space-time structure
of the matrix configurations obtained 
in Monte Carlo studies
of
%by simulating 
the Lorentzian type IIB matrix model and the simplified models.
In these models, the time-evolution can be extracted 
from the matrix configurations
by working in the SU($N$) basis which diagonalizes the temporal matrix $A_0$.
The $n\times n$ spatial submatrices 
$\bar{A}_i(t)$ ($i=1,\cdots ,9$)
at each time $t$ show that only three out of nine directions expand
% with $t$
after some critical time suggesting
the SSB of rotational SO(9) symmetry to SO(3).
By calculating the eigenvalues of $\bar{A}_i(t)$ at each $t$, however, 
we have found that only two of them increase in magnitude with $t$ 
in the extended directions, while the rest are independent of $t$ 
and SO(9) symmetric.
This implies that the SSB is caused only by the two eigenvalues.
% at each time.
In the continuum limit, the magnitude of the two eigenvalues diverges
in physical units
% at each time 
and the spatial matrices $\bar{A}_i(t)$
approach a configuration which is essentially 
described by the Pauli matrices.
%% This Pauli-matrix structure is quite robust, and it does not seem to be
%% avoided even in the original supersymmetric model 
%% including fermionic contributions.

We have attributed this problem
to the approximation used in Monte Carlo simulation 
to avoid the sign problem,
which actually amounts to replacing
$\ee^{iS_{\mathrm{b}}}$ by 
$\ee^{\beta S_{\mathrm{b}}}$
%$\delta(S_{\mathrm{b}})$ 
in the partition function
(\ref{eq:partition-function}) of the Lorentzian type IIB matrix model.
This new interpretation of the Monte Carlo simulation
enables us to understand the interesting aspects of the obtained results
such as the band-diagonal structure of the spatial matrices $A_i$
%in the SU($N$) basis which diagonalizes $A_0$ 
as well as the appearance of 
the (3+1)d expanding behavior
with the Pauli-matrix structure.
%based on some classical analysis.

In order to discuss what happens in the original model,
we have considered a model with a factor
$\ee^{\beta (\cos\theta  + i \sin \theta) S_{\rm b}}$,
%with $0 \le \theta \le \pi/2$, 
which interpolates the model
we have been studying ($\theta=0$)
and the model we are aiming at ($\theta=\pi/2$).
Using some arguments based on the classical equation of motion,
which is common to all $\theta$,
we have speculated that it is possible to obtain a regular space-time
structure with the (3+1)d expanding behavior 
by approaching $\theta=\pi/2$ in the large-$N$ limit.
The crucial point is that the Pauli-matrix structure is
obtained by maximizing the action at the expense of reducing the
entropy. By approaching $\theta=\pi/2$, one may obtain
classical solutions which only extremize the action that
have larger entropy due to a smooth eigenvalue distribution of the
matrix $Q(t)$. The existence of such classical solutions 
with the (3+1)d expanding behavior 
has been confirmed numerically \cite{classical}.
%Whether classical solutions with the (3+1)d expanding behavior 
%
Whether such classical solutions appear from the full quantum
theory by approaching $\theta=\pi/2$
remains to be seen.
%% Our argument suggests
%% that such a configuration may play
%% the role of a seed for a regular (3+1)d expanding space-time
%% in the Lorentzian type IIB matrix model.
%SSB of SO(9) rotational symmetry
%to SO(3) in the Lorentzian type IIB matrix model.

Monte Carlo simulation of the interpolating model for $\theta \neq 0$
is difficult since
the complex
weight $\ee^{\beta (\cos\theta  + i \sin \theta) S_{\rm b}}$ 
causes the sign problem.
As a promising approach to overcome this problem,
we may use the complex Langevin method \cite{Parisi:1984cs,Klauder:1983sp}, 
which has attracted much attention
recently \cite{Aarts:2009dg,Aarts:2009uq,Aarts:2011ax,Nishimura:2015pba,Nagata:2015uga,Nagata:2016vkn,Ito:2016efb}.
It was successful also 
in investigating
the SSB of rotational symmetry in
the 6d Euclidean type IIB matrix model \cite{Anagnostopoulos:2017gos}.
%% In fact, after an appropriate change of variables,
%% it turns out that the complex Langevin method works
%% % for $\theta \neq 0$
%% at least in the bosonic model.
Preliminary results \cite{workinprog} for the bosonic Lorentzian model
suggest that by approaching $\theta = \pi/2$, 
one obtains clear deviations from the Pauli-matrix structure
without losing the (3+1)d expanding behavior.
We hope to see whether a regular (3+1)d expanding space-time
emerges or not
%appears in the along this line 
in the near future.

%% while the other important properties such as
%% the band-diagonal structure of $A_i$ and the expanding (3+1)d behavior
%% are kept intact.

%% Last but not the least, the structure in the extra dimensions
%% is also important since it determines the low energy effective theory
%% in the (3+1)d space-time.
%% We consider it possible 
%% that the extra dimensions have a rich structure
%% by approaching $\theta = \pi/2$ at large $N$.

%% We hope to see whether
%% this seed turns into a matrix configuration
%% representing a regular continuous 3d space in 
%% an appropriate large-$N$ limit.
%% In that limit, it is also expected that the extra dimensions
%% acquire richer structure than what is observed for $\theta=0$.

%is an important question we would like to answer 
%in the near future.
%% We hope to see how this seed 
%% establish the emergence of a regular space-time 
%% with the (3+1)d expanding behavior in the near future.

%%%%%%%%%%%%%%%%%%%%%%%%%%%%%%%%%%%%%%%%%%%%%%%%%%%%%%%%%%%%%%%%%%%%%%
%%%%%%%%%%%%%%%%%%%%%%%%%%%%%%%%%%%%%%%%%%%%%%%%%%%%%%%%%%%%%%%%%%%%%% 

%%%%%%%%%%%%%%%%%%%%%%%%%%%%%%%%%%%%%%%%%%%%%%%%%%%%%%%%%%%%%%%%%%%%%%
%%%%%%%%%%%%%%%%%%%%%%%%%%%%%%%%%%%%%%%%%%%%%%%%%%%%%%%%%%%%%%%%%%%%%%
%%%%%%%%%%%%%%%%%%%%%%%%%%%%%%%%%%%%%%%%%%%%%%%%%%%%%%%%%%%%%%%%%%%%%%
\section*{Acknowledgements}
%\hspace{0.51cm}
%%%%%%%%%%%%%%%%%%%%%%%%%%%%%%%%%%%%%%%%%%%%%%%%%%%%%%%%%%%%%%%%%%%%%%
%%%%%%%%%%%%%%%%%%%%%%%%%%%%%%%%%%%%%%%%%%%%%%%%%%%%%%%%%%%%%%%%%%%%%%
%%%%%%%%%%%%%%%%%%%%%%%%%%%%%%%%%%%%%%%%%%%%%%%%%%%%%%%%%%%%%%%%%%%%%% 

The authors would like to thank 
K.N.~Anagnostopoulos, T.~Azuma,
K.~Hatakeyama,
S.~Iso, H.~Kawai,
A.~Matsumoto,
H.~Steinacker 
and A.~Yosprakob for valuable discussions.
This research used computational resources 
of the K computer of the HPCI system 
provided by the AICS 
%through the HPCI System Research Project (Project ID : hp130063).
through the HPCI System Research Project 
``Quest for the ultimate laws and the birth of the universe''
(Project ID : hp170229, hp180178).
Computation was carried out also on PC clusters at KEK.
The supercomputer FX10 at
University of Tokyo has been used 
in developing our code for parallel computing.
A.~T.\ was supported in part by Grant-in-Aid 
for Scientific Research (No.\ 18K03614)
from Japan Society for the Promotion of Science. 

%% J.~N.\ was supported in part by Grant-in-Aid 
%% for Scientific Research (No.\ 16H03988)
%% from Japan Society for the Promotion of Science. 

\appendix

\section{The determination of the parameter $p$}
\label{sec:appendix_universality} 
%%%%%%%%%%%%%%%%%%%%%%%%%%%%%%%%%%%%%%%%%%%%%%%%%%%%%%

In this appendix, we explain how we determine the parameter $p$
in the IR cutoff (\ref{eq:cutoff-temp}) and 
(\ref{eq:cutoff-space}).
While a naive choice would be $p=1$,
it was proposed in ref.~\cite{Ito:2017rcr} that
one should choose a slightly larger value so that
the results become almost independent of $p$.
There it was found in the VDM model
that the results for the extent of space $R^2(t)$ 
become independent of $p$
when $p$ is larger\footnote{For the values of $p$ in this region, 
it was also observed \cite{Ito:2017rcr} from the analysis of the 
Schwinger-Dyson equations
that the effect of the IR cutoff decreases as one takes the 
infinite volume limit.} than $p_{\rm c}=1.2 \sim 1.3$.
Based on this observation,
we used $p=1.4$ when we simulate the VDM model 
in section \ref{sec:cont-lim}.

\begin{figure}
\centering{}
\includegraphics[scale=0.6]{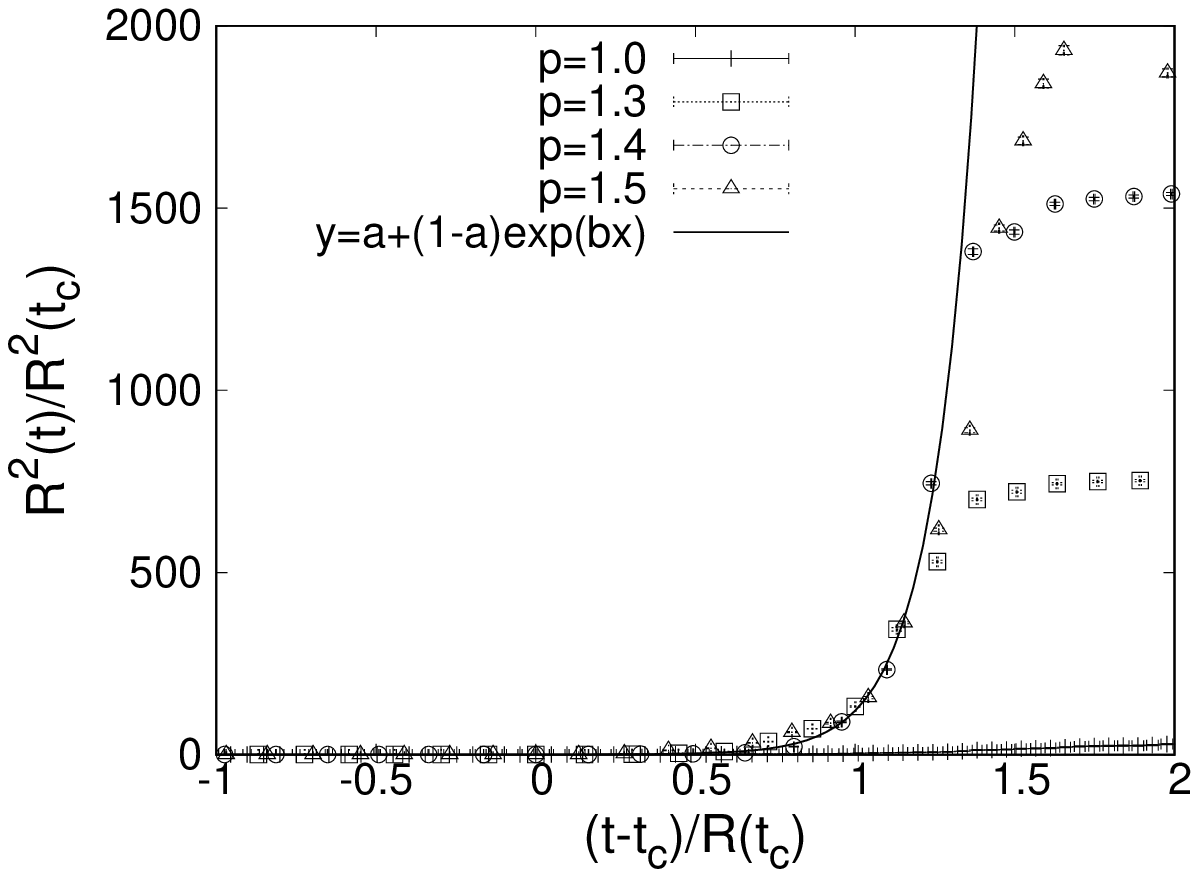}
\includegraphics[scale=0.6]{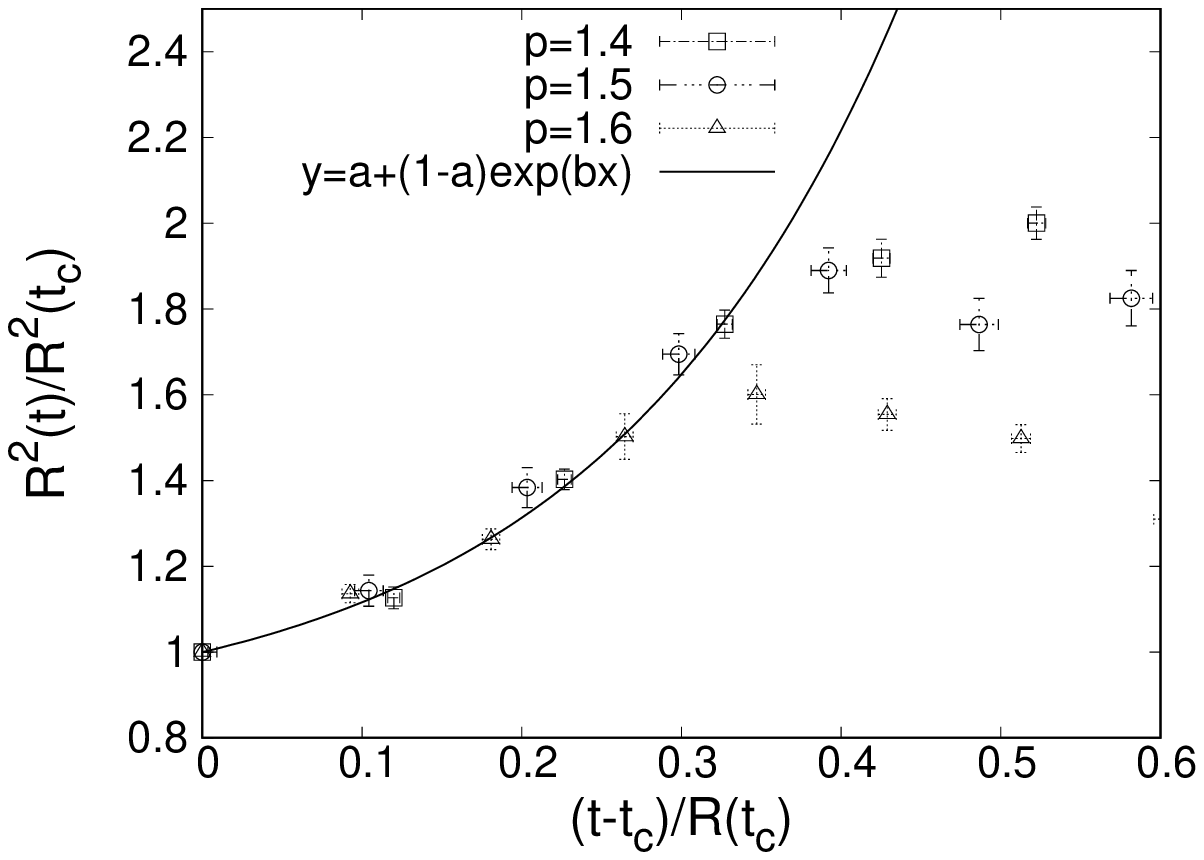}
\caption{(Left) The extent of space $R^{2}(t)/R^{2}(t_{\rm c})$ 
obtained for the bosonic model
is plotted against $x=(t-t_{\rm c})/R(t_{\rm c})$ 
for various values of $p$ with $N=256$, $C=100$, $\kappa=1.0$. 
The block size is chosen as $n=32, 24, 20, 18$ for
$p=1.0, 1.3, 1.4, 1.5$, respectively.
The solid line represents 
a fit to the $p=1.4$ data with $R^{2}(t)/R^{2}(t_{\rm c})=
a+(1-a)\exp(bx)$, which gives $a=0.92(5)$, $b=7.3(6)$.
%
%a=0.9242(549)
%b=7.3362(6477)
%
(Right) The extent of space $R^{2}(t)/R^{2}(t_{\rm c})$ 
obtained for the original model
is plotted
against $x=(t-t_{\rm c})/R(t_{\rm c})$ 
for various values of $p$ with $N=16$,
$C=5$, $\kappa=0.46$. 
The block size is chosen as $n=7, 6, 6$ for
$p=1.4, 1.5, 1.6$, respectively.
The solid line represents a fit to the $p=1.6$ data 
with $R^{2}(t)/R^{2}(t_{\rm c})=a+(1-a)\exp(bx)$,
which gives $a=0.83(4)$, $b=5.3(7)$.
%a=0.8337(423)
%b=5.2948(6914)
}
\label{univ_bos|univ_ikkt} 
\end{figure}

Here we repeat 
the same analysis in the case of 
the bosonic model and the original model.
%% in sections 
%% \ref{sec:bosonic} and \ref{sec:fermions}.
%
%to determine a suitable value of $p$ to be used
%
%in order to determine the suitable value for the parameter $p$
%in these models, which should be used 
In Fig.~\ref{univ_bos|univ_ikkt}, we plot the extent of space 
$R^{2}(t)/R^{2}(t_{\rm c})$ 
against time 
$(t-t_{\rm c})/R(t_{\rm c})$ 
%$t$ 
for the bosonic model (Left)
and the original model (Right), respectively,
with various values of $p$.
For all values of $p$,
we find that only three directions start to expand 
at some critical time $t_{\rm c}$.
In the bosonic model, the results 
%for $t>t_{\mathrm{c}}$
scale for $p=1.3, 1.4, 1.5$ 
except for the data around the peak of $R^{2}(t)$.
%whereas the results for $p=1.0$ do not lie on the scaling curve. 
Similar scaling behavior is observed
for the original model
for $p=1.4, 1.5, 1.6$.
Based on these results, we use $p=1.5$ for the bosonic model
and $p=1.6$ for the original model
in sections \ref{sec:bosonic} and \ref{sec:fermions}, respectively.

\bibliographystyle{JHEP}
\bibliography{ref}

%\bibliographystyle{h-physrev5}
%\bibliographystyle{JHEP}
%\bibliographystyle{utphys}
%\bibliography{ref}

%%%%%%%%%%%%%%%%%%%%%%%%%%%%%%%%%%%%%%%%%%%%%%%%%%%%%%%%%

\end{document}